\documentclass[manuscript,screen]{acmart}

\usepackage{xcolor}
\usepackage{bbm}

\newcommand{\highlight}[1]{\textcolor{black}{#1}}

\AtBeginDocument{%
  }

\setcopyright{acmlicensed}
\copyrightyear{2018}
\acmYear{2018}
\acmDOI{XXXXXXX.XXXXXXX}

\acmConference[Conference acronym 'XX]{Make sure to enter the correct
  conference title from your rights confirmation emai}{June 03--05,
  2018}{Woodstock, NY}
\acmISBN{978-1-4503-XXXX-X/18/06}

\begin{document}

\title{MV-Crafter: An Intelligent System for Music-guided Video Generation}

\author{Chuer Chen}
\email{chuerchen1998@gmail.com}
\affiliation{%
  \institution{Tongji University}
  \country{China}
}

\author{Shengqi Dang}
\affiliation{%
  \institution{Tongji University}
  \country{China}}
\email{larst@affiliation.org}

\author{Yuqi Liu}
\affiliation{%
  \institution{Tongji University}
  \country{China}
}
\email{larst@affiliation.org}

\author{Nanxuan Zhao}
\affiliation{%
 \institution{Adobe Research}
 \country{USA}}
\email{nanxuanzhao@gmail.com}

\author{Yang Shi}
\affiliation{%
  \institution{Tongji University}
  \country{China}}
\email{yangshi.idvx@tongji.edu.cn}

\author{Nan Cao}
\authornotemark[2]
\authornote{Nan Cao is the corresponding author.}
\affiliation{%
  \institution{Tongji University}
  \country{China}}
\email{nan.cao@tongji.edu.cn}

\renewcommand{\shortauthors}{Chen et al.}

\begin{abstract}
    Music videos, as a prevalent form of multimedia entertainment, deliver engaging audio-visual experiences to audiences and have gained immense popularity among singers and fans. Creators can express their interpretations of music naturally through visual elements. However, the creation process of music video demands proficiency in script design, video shooting, and music-video synchronization, posing significant challenges for non-professionals. Previous work has designed automated music video generation frameworks. However, they suffer from complexity in input and poor output quality. In response, we present MV-Crafter, a system capable of producing high-quality music videos with synchronized music-video rhythm and style. Our approach involves three technical modules that simulate the human creation process: the script generation module, video generation module, and music-video synchronization module. MV-Crafter leverages a large language model to generate scripts considering the musical semantics. To address the challenge of synchronizing short video clips with music of varying lengths, we propose a dynamic beat matching algorithm and visual envelope-induced warping method to ensure precise, monotonic music-video synchronization. Besides, we design a user-friendly interface to simplify the creation process with intuitive editing features. Extensive experiments have demonstrated that MV-Crafter provides an effective solution for improving the quality of generated music videos.
\end{abstract}




\received{20 February 2007}
\received[revised]{12 March 2009}
\received[accepted]{5 June 2009}

\maketitle

\section{Introduction}
Music video, a type of media combining both music and visual content, creates a multidimensional experience for audiences. With social media's power, the popularity of music videos is significantly boosted. A music video can promote the music by inducing audiences' emotions and reactions to improve the impression, understanding, and memorability. To produce a compelling music video, creators often follow a three-step process~\cite{robin2022howtomakemv,robinhowtoplanemv,anna2023howtomakemv,ronfard2021film}: pre-production: drafting scripts/storyboard to conceptualize the video content based on the music; production: collecting video resources by shooting or retrieving existing repositories; post-production: composing for synchronization with editing tools to obtain the final music video. This can be a long process, which is time-consuming and requires professional experience.

Many research works have appeared to study the problem of easing this process, which can be categorized into two categories: retrieval-based methods~\cite{emv-matchmaker,Lin2017AutoMV,liao2015audeosynth,Yoon2009AutomatedMV, suris2022s} and generation-based methods~\cite{Liu2023GenerativeDT, Jeong2021TrumerAIDM,Jang2022Music2VideoAG}. Retrieval-based methods retrieve videos from user-collected repositories based on the given music or textual input, while generation-based methods directly predict music video with generative models such as Generative Adversarial Networks (GAN)~\cite{Jang2022Music2VideoAG,Jeong2021TrumerAIDM} and diffusion models~\cite{Liu2023GenerativeDT}. However, creating high-quality music videos with these methods still faces three challenges: 1) Lack of narrative. Previous methods~\cite{match-music-image, Jang2022Music2VideoAG} retrieve mainly based on lyrics, which are abstract and discrete, lacking coherent semantic guidance. 2) Low visual quality and diversity. Since retrieval-based methods obtain the clips from pre-defined limited repositories, the diversity is highly restricted. Though the generation-based method can generate arbitrary content, the visual quality is not guaranteed. 3) Inaccurate synchronization. Music rhythms are complex and expressive, increasing the difficulties in matching up with video clips under various lengths. How to smoothly transit among video clips while matching with audio beats is a difficult problem. 

To this end, we propose an intelligent music video generation system called \textbf{MV-Crafter} that creates synchronized videos based on the input music and themes (i.e., a few keywords describing the desired content of the music video). The system mimics the creator operations and consists of three modules: 1) Script generation module: After segmenting the input music into clips based on the rhythm, we take a three-step strategy to generate scripts (i.e., a series of prompts, with each one describing a scene) and keywords based on theme and music clips. By taking advantage of LLM~\cite{openai2023gpt4} and pre-trained music captioning model~\cite{Doh_Choi_Lee_Nam_2023}, our module can generate more accurate scripts that seamlessly blend with the musical semantics. 2) Video generation module: It utilizes powerful text-to-image and image-to-video diffusion models~\cite{stablediffusion,stablevideodiffusion} to generate video clips given the predicted scripts and keywords, streamlining the laborious preparation of visual elements by creators. 3) Dynamic music-video synchronization module: To rhythmically align the music and video clips, our core idea is innovated by a beat-matching algorithm for key frames using dynamic programming, and a visual envelope-induced warping algorithm for the remaining frames to form an accurate and monotonic synchronization.

We design a user-friendly interface for creators to easily use our system. Users can simply upload music, input a theme, and gradually generate a music video through clicks and text interactions. Our experiments demonstrate the system's effectiveness in generating high-quality, rhythmic music videos. 
% In case users are dissatisfied with the generated results, adjustments can be made by modifying the script content or replacing scene images.
Generally, we make the following contributions:
\begin{itemize}
  \item We present an interactive system and novel generation pipeline that enables non-professionals to create rhythmic music videos via an input piece of music and theme. 
  \item We propose a dynamic music-video synchronization technique, which aligns video clips with music clips of arbitrary lengths while ensuring the monotonicity of the warping curve.
  \item We introduce a novel music video script generation technique that produces a script as a sequence of consecutive prompts that meaningfully matches the style and the topics of the input music and theme. These prompts are later used for generating video clips of the music video.
\end{itemize}

\section{Related Works}
In this section, we discuss prior works that are closely related to our method, namely, music video generation, audio-visual synchronization, and script-related video generation.

\subsection{Music Video Generation}
Traditional techniques generate music videos by concatenating images retrieved through lyrics~\cite{musicstory, match-music-image, Cai2007AutomatedMV}. With advancements in video processing techniques, the mainstream has shifted towards matching music and video~\cite{Hua2004AutomaticMV, Shin2016AutomatedMV, emv-matchmaker, Fan2016DJMVPAA, Gross2019AutomaticRM, Lin2016AutomaticMV, Lin2017AutoMV, suris2022s}. Some studies focus on the perceptual connection between music and video that design emotion-driven matching methods~\cite{Shin2016AutomatedMV, emv-matchmaker, Fan2016DJMVPAA, Lin2016AutomaticMV, Lin2017AutoMV, liu2023emotionaware}. For example, 
% EMV-matchmaker~\cite{emv-matchmaker} adopt an emotional temporal course model (ETCM) to predict the emotional temporal phase sequence of video (or music). 
Liu et al.~\cite{liu2023emotionaware} propose a two-stage framework for movie montage generation, including a learning-based module for predicting emotion similarity and an optimization-based module for selecting and composing movie shots. Additionally, some approaches involve retrieving music or videos based on extracted low-level features like velocity~\cite{Yoon2009AutomatedMV}, saliency~\cite{liao2015audeosynth}, and color features~\cite{Gross2019AutomaticRM, Fan2016DJMVPAA}. However, the retrieval methods restrict the diversity of video content. Alternatively, most works emphasize the matching of music and video, with few achieving rhythm synchronization~\cite{Gross2019AutomaticRM, emv-matchmaker, Lin2016AutomaticMV, Shin2016AutomatedMV} or relying on simple methods like duration constraints~\cite{liu2023emotionaware} or dynamic time warping~\cite{Lin2017AutoMV}.

Recently, generative AI has provided a broader creative space for music video generation. Tr\"{a}umerAI~\cite{Jeong2021TrumerAIDM}  employs StyleGAN to visualize music by mapping music embeddings to style embeddings. Music2Video~\cite{Jang2022Music2VideoAG} utilizes Wav2CLIP~\cite{wu2022wav2clip} and CLIP~\cite{radford2021learning} to encode music and lyrics into latent representations to guide VQ-GAN~\cite{esser2021taming} in generating videos. Generative Disco~\cite{Liu2023GenerativeDT} introduces Stable Diffusion Videos~\cite{stablediffusionvideo} and GPT-4~\cite{openai2023gpt4} to generate stylized music videos from input prompts and music. However, their generated music videos often suffer from poor visual quality and lack of coherent content. In contrast, MV-Crafter is able to generate high-quality, coherent, and rhythmic music videos.

\subsection{Audio-Visual Synchronization}
Audio-visual synchronization is a pivotal technique within multimedia production that involves aligning audio and visual elements to ensure consistency between them.
% Audio-visual synchronization is a crucial aspect of music video production, involving the alignment of audio and visual elements to ensure consistency between them.
Research in this domain has explored dance-music ~\cite{AlignNet,zhou2023let,bellini2018dance,yu2022self} and lip-speech ~\cite{AlignNet,halperin2019dynamic,chung2017out} synchronization. Wang et al.~\cite{AlignNet} propose AlignNet, an end-to-end trainable model that learns the mapping between visual dynamics and audio rhythms, requiring training on task-specific datasets for respective tasks. For open-domain synchronization tasks, some works synchronize videos with target audio signals through beats alignment and temporal warping~\cite{chen2011visual,liao2015audeosynth, davis2018visual,sun2023eventfulness}. A notable contribution in this domain is Visbeat~\cite{davis2018visual}, which proposes a flow-based analysis method to quantify the rhythm of a video. This method aligns extracted visual beats with musical beats through time-warping based on an unfolding strategy. Furthermore, Sun et al.~\cite{sun2023eventfulness} introduce visual eventfulness as an indicator of significant points that affect visual perception within videos. They construct a synthetic video dataset containing noticeable visual impacts and train a deep neural network to predict the eventfulness of the videos. 

Considering the specificity of AI-generated videos, we opted for the rule-based VisBeat approach for visual rhythm extraction due to its higher generalization capabilities. Additionally, we observed that the warping methods applied in previous works could result in repetitive video frames when the target music is longer, making them unsuitable for music videos. To address this, we propose a novel synchronization method for music videos aimed at ensuring consistency in rhythm and temporal monotonicity.
% Therefore, we propose a synchronization method that ensures both rhythmic consistency and monotonicity in time.

\subsection{Script-Related Video Generation}
Script-related video generation is a specialized form of Text-to-Video (T2V) generation that focuses on generating videos from coherent and multi-scene textual prompts. Recent works~\cite{luo2023videofusion,wu2023tune,khachatryan2023text2video,zhang2023show} in T2V rely on diffusion models for its excellent performance in producing high-quality videos. Some T2V tasks shift focus to creating videos with storylines and multiple scenes, utilizing large language models (LLM) for script writing. Animate-A-Story~\cite{he2023animate}, MovieFactory~\cite{zhu2023moviefactory} and TaleCrafter~\cite{gong2023talecrafter} expand the input text into a script that includes a sequence of prompts for image/video generation. VideoDrafter~\cite{long2024videodrafter} utilizes LLM to extract entities from the generated script to maintain entity consistency in generated videos. Vlogger~\cite{Zhuang2024vlogger:} employs an iterative dialogue approach for script creation and actor design.
The script generation methods in the aforementioned works all generate the script from the story theme. \highlight{To the best of our knowledge, we are the first to propose a music video script generation method that enhances the script with musical semantics.} 
% Our script generation method enhances the script with music semantics and automatically generates style keywords to ensure consistent style across different scenes.

\section{System Overview}
Inspired by the rapid development of video generation techniques, we design MV-Crafter, an intelligent music video generation system leveraging AI technology to streamline the creative process. In this section, we first summarize the design requirements according to a preliminary survey. After that, we introduce the architectural design of MV-Crafter. Finally, we present the system's user interface and functionalities.

\subsection{Design Requirements}
To design a music video generation and authoring system, it is necessary to understand how a typical music video is created and the limitations of existing techniques. To this end, we reviewed a number of tutorials about how to produce music videos written by music video directors ~\cite{robin2022howtomakemv,robinhowtoplanemv, anna2023howtomakemv}. All of them suggest a process consisting of three key steps: (1) pre-production, in which directors conceptualize the music and draft a storyboard containing visual narratives and shots; (2) production, in which directors shoot the video footage based on the storyboard; and (3) post-production, in which directors synchronize the video footage with music via authoring tools (e.g., Adobe Premiere). 

Following the above procedure, several intelligent music video generation tools such as Kaiber~\cite{kaiber}, Plazmapunk~\cite{plazmapunk} have been developed recently based on generative AI models. These tools take prompts as input to generate a series of consecutive video clips. Users need to write prompts, manually splice the video clips, and synchronize them to the music. The resulting music videos largely fall into two styles: (1) continuous motion videos (e.g., videos generated by Pika~\cite{pika} and Gen-2~\cite{gen2}) and (2) frame-by-frame animation with audio reactivity (e.g., videos generated by Kaiber~\cite{kaiber}).

The above survey suggested that although AI techniques greatly lower the barrier of video production, prompt writing and music-video synchronization are the crucial steps that affect the quality of the generated music videos, but still take users' efforts. Therefore, our goal is to design an intelligent system that not only generates videos but also helps with story prompt generation and automatic music-video synchronization. We elaborate this goal on the following design requirements: 

\begin{enumerate}
\item[{\bf R1}]{\bf Generating narrative scripts aligned with music and theme.} It usually takes effort to understand the music and design scenes when crafting storyboards. The system should automatically generate narrative scripts aligned with the user's input theme (i.e., a few keywords describing the potential topic) and music. These scripts should be able to be used as prompts describing each scene for subsequent video generation.

\item[{\bf R2}] {\bf Generating smoothly animated videos.} The system should generate video clips with continuous motions based on the scripts to visually narrate the storyboard smoothly.

\item[{\bf R3}] {\bf Automatic music-video synchronization.} Synthesizing rhythmic music videos entails both time investment and proficiency in video editing, along with a deep understanding of musical structure. Therefore, the system should automatically splice the generated video clips with the music based on the "cut-to-the-beat" rule and synchronize the visual and musical rhythms.

\item[{\bf R4}] {\bf Flexible editing on the generated results.} Using the system, users should have the ability to edit all generated results throughout the pipeline, including scripts, scene images, and videos.

\end{enumerate}

\begin{figure*}[!htb]
    \centering
    \includegraphics[width=\linewidth]{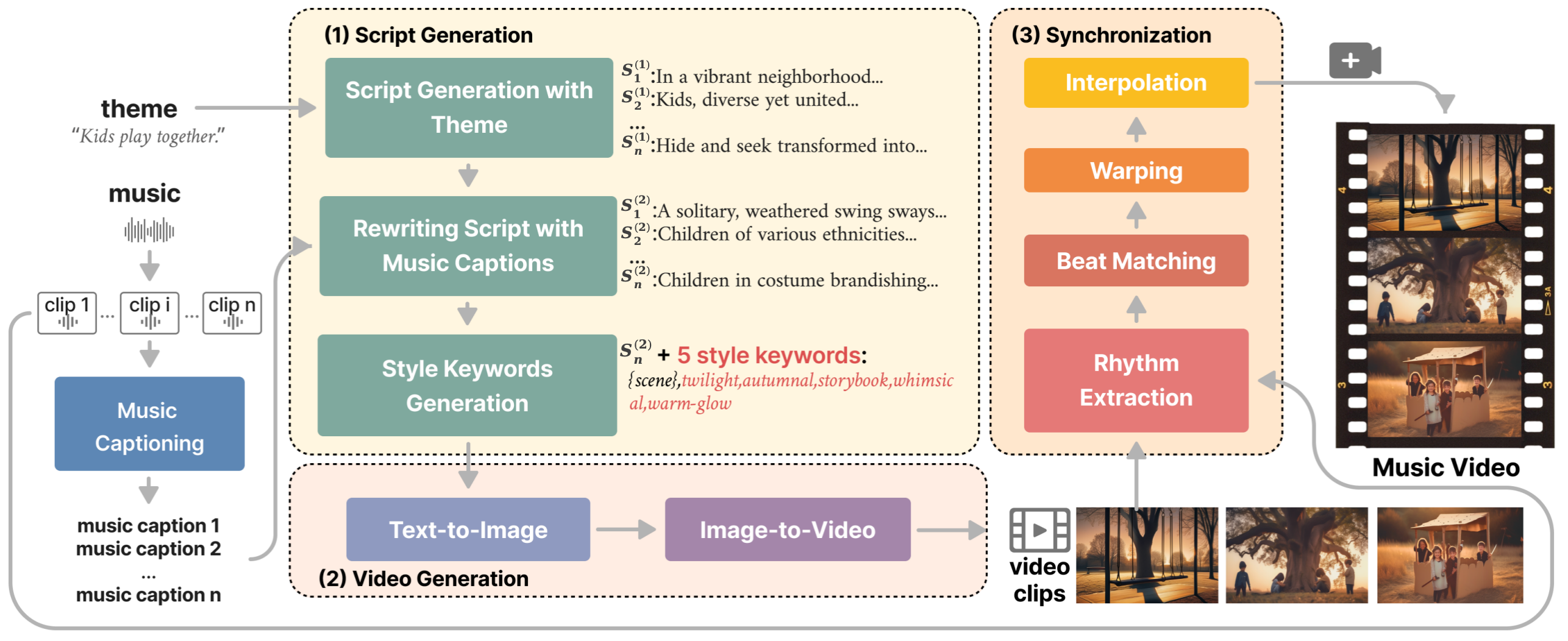}
    \caption{The pipeline of MV-Crafter consists of three modules: (1) a script generation module for generating music video scripts and keywords; (2) a video generation module for creating video clips;(3) a synchronization module dedicated to aligning the rhythm between music and video clips.}
    \label{fig:overview}
    % \vspace{-1.5ex}
\end{figure*}

\subsection{System Design}
To fulfill these design requirements, we introduce MV-Crafter, an intelligent and interactive system to assist non-professionals in creating insightful and rhythmic music videos via input music and theme. Fig.~\ref{fig:overview} illustrates the architecture and pipeline of the proposed MV-Crafter system. Generally, it consists of three major modules: (1) the script generation module, (2) the video generation module, and (3) the synchronization module. Given input music and the corresponding theme describing the desired music video content, we first split the music into clips by its rhythms. A caption describing the music (e.g. genre, instrument, mood) is generated for each music clip based on a pre-trained music captioning model~\cite{Doh_Choi_Lee_Nam_2023}. The script generation module takes both the theme and music captions as input to produce a script that considers the semantics of the music via GPT-4~\cite{openai2023gpt4} ({\bf R1}). The video generation module takes the script (i.e., prompts) as the input to generate static images that drive an image-to-video generation model~\cite{stablevideodiffusion} to generate video clips corresponding to each music clip ({\bf R2}). Finally, the synchronization module aligns the music clips and the video clips by optimally matching the music beats and visual beats, warping the remaining video frames ({\bf R3}), and then concatenating the results to generate rhythmical music videos.

\begin{figure*}
  \centering
  \includegraphics[width=\textwidth]{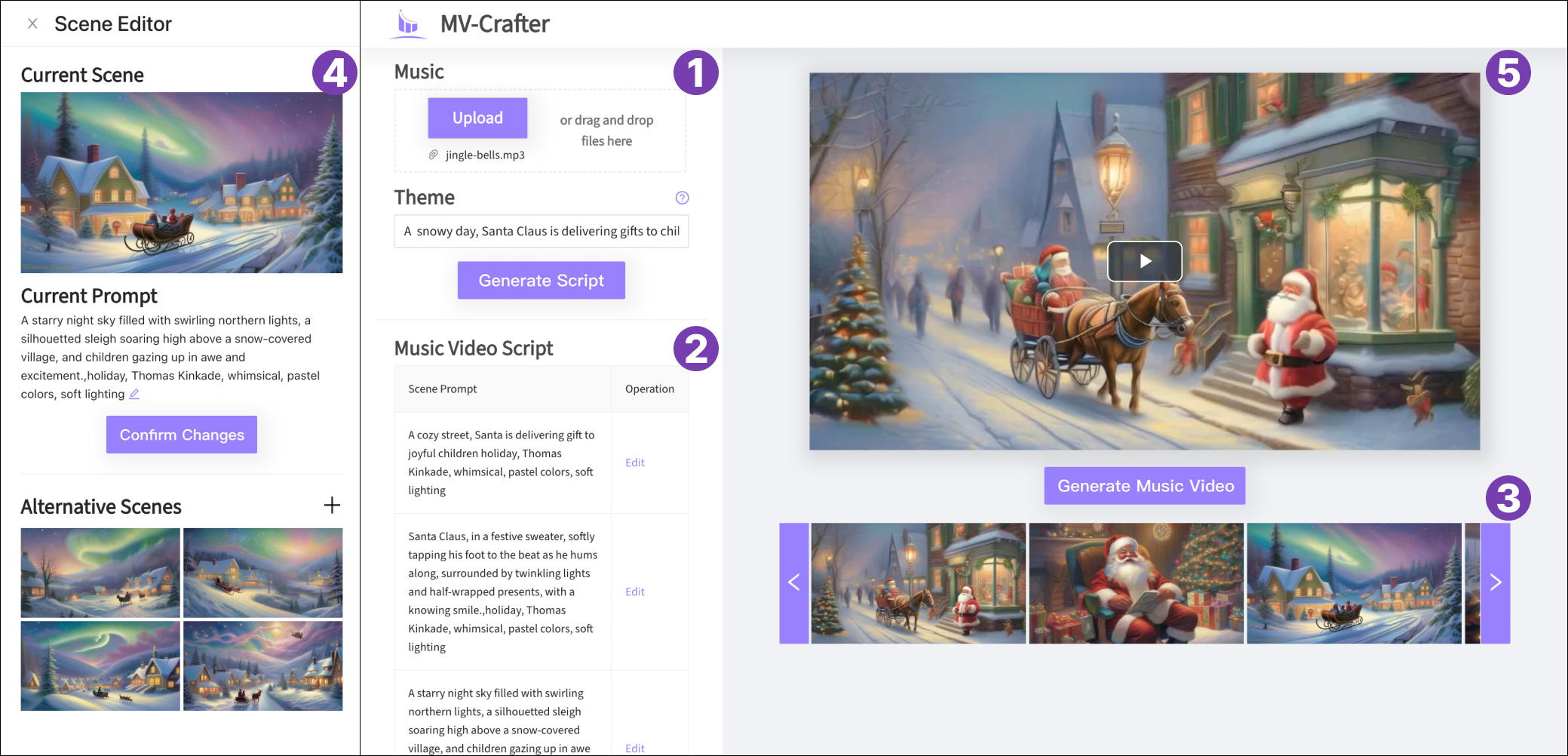}
  \caption{The user interface of MV-Crafter. The right panel displays the main interactive interface of the system. The left panel is the scene editor, where users can edit and replace individual scenes in this pop-up window.}
  \label{fig:interface}
\end{figure*}

\subsection{User Interface.} 
We design a user interface (Fig.~\ref{fig:interface}) for the system to help users generate and edit music videos ({\bf R4}). Specifically, after the user uploads a music file and enters the theme of the desired music video content in the input field (Fig.~\ref{fig:interface}(1)), the music video script will be generated scene by scene as displayed in Fig.~\ref{fig:interface}(2), which can also be modified by users. The scene images generated based on the script are shown as the thumbnails of the scenes in the timeline window(Fig.~\ref{fig:interface}(3)). By clicking a thumbnail, the user can further modify the scene prompt, generate alternative scenes, and select a satisfactory scene in a popup window (Fig.~\ref{fig:interface}(4)). Once everything is done, a music video can be automatically generated by clicking the generation button. The resulting music video will be played in the playback window (Fig.~\ref{fig:interface}(5)). If any scene fails to meet expectations, users can iteratively edit it until fulfilling their requirements.

\section{Music Video Generation Pipeline}
In this section, we introduce the preprocessing method for handling input music, followed by detailed descriptions of the script generation module, video generation module, and dynamic music-video synchronization module.

\subsection{Preprocessing}
Music videos are usually multi-scenes long videos, where each video scene has a rhythmic and semantic connection to the corresponding music clip. To provide music clips as conditions for generating video scenes, we first segment the uploaded music based on its beats by iteratively identifying local maxima in the onset strength of beats as segmentation points until the average duration of music clips approximates the bar duration $\Delta_{bar}=4\Delta_{beat}=4\frac{60}{bpm}$. This segmentation method meets two criteria: 1. According to the "cut-to-the-beat" rule~\cite{vernallis2004experiencing}, segmentation points should coincide with strong beats. 2. The duration of video scenes is generally associated with the bar duration~\cite{prétet2021language}.

Later, we apply a transformer-based music captioning model, LP-MusicCaps~\cite{Doh_Choi_Lee_Nam_2023}, to generate captions for the segmented music clips. Each caption describes the genre, instrument, vocal, mood, tempo, culture, and sonority features of the corresponding music clip, providing additional semantic context about the music.

% We iteratively identify local maxima in the music beats sequence as segment points until the average duration of the segments exceeds the bar duration. 

\subsection{Music Video Script Generation} \label{script-generation}
In this module, we input both the user-inputted theme and captions of $N$ music clips into a Large Language Model (LLM) to generate a script infused with musical semantics. This process yields a script consisting of $N$ scene prompts, with each prompt describing a scene, and also produces five style keywords to ensure the stylistic coherence of the resulting music video. Recognizing the challenge faced by LLMs in capturing long-range context, we design a three-step prompting method to gradually guide the LLM. Illustrated in Fig.~\ref{fig:overview}(1), the process comprises three main steps: (1) Expanding the user-inputted theme into a script with multiple scenes; (2) Rewriting the script by incorporating music captions; (3) Generating style keywords for the script to derive the final music video script.

\textbf{Script Generation with Theme.}
The user-inputted theme $T$ may encompass elements such as characters, plot, and setting. By leveraging the robust generalization capabilities of LLM, we can expand a simple theme into a coherent story. Specifically, we utilize the LLM to generate scene prompts equal in number to the music clips. The process can be defined as follows:

% \begin{equation}
%     [S_1,S_2,...,S_n]=LLM(I_1,T,N)
% \end{equation}
\begin{equation}
    \{S^{(1)}_1,S^{(1)}_2,...,S^{(1)}_N\}=LLM(I^{(1)},T,N),
\end{equation}
% \begin{equation}
%     \left\{S^{(1)}_i\right\}=LLM(I^{(1)},T,N)
% \end{equation}
where $I^{(1)}$ represents the instruction of script generation given to LLM. $\{S^{(1)}_i\}$ contains the $N$ prompts of scenes in the script. $N$ is the number of music clips. 

\textbf{Rewriting Script with Music Captions.}
To match the video content with the musical style, we reference Music-To-Image~\cite{music-to-image} to guide the script generation process with music captions. We task LLM with rewriting generated script $S^{(1)}$, incorporating corresponding music captions $M$. 
% The requirement is for the rewritten script to integrate with both the preceding script and the semantic content of the music. This process can be formally defined as follows:

% \begin{equation}
%     R_i=LLM(I_2,S_i,M_i),i=1,2,...,n
% \end{equation}
\begin{equation}
    S^{(2)}_i=LLM(I^{(2)},S^{(1)}_i,M_i),
\end{equation}
where $I^{(2)}$ denotes the instruction of rewriting provided to the LLM. $M_i$ corresponds to the music caption of the $i$-th music clip, while $S^{(2)}_i$ represents the prompt of the $i$-th scene in the rewritten script that integrates the original scene prompt with the music caption.

\textbf{Style Keywords Generation.}
Upon generating the script incorporating music captions, 
% LLM is instructed to generate 5 style keywords based on the all rewritten script. 
LLM is instructed to generate five visual style keywords that align with the overall scene in the script $S^{(2)}$.

Let $K$ denote the set of five style keywords, $I^{(3)}$ denotes the instruction for keywords generating related to visual style. The generation process can be defined as
% \begin{equation}
%     K=LLM(I_3,[R_1,R_2,...,R_n])
% \end{equation}

\begin{equation}
    K=LLM(I^{(3)},\{S^{(2)}_1,S^{(2)}_2,...,S^{(2)}_N\}).
\end{equation}

To maintain stylistic consistency throughout various scenes in the music video, the keywords are then concatenated with each scene of the rewritten script, forming a complete music video script $S$,
\begin{equation}
    S=\{S^{(2)}_1+K,S^{(2)}_2+K,...,S^{(2)}_N+K\}.
\end{equation}

\begin{figure}
    \centering
    \includegraphics[width=0.65\textwidth]{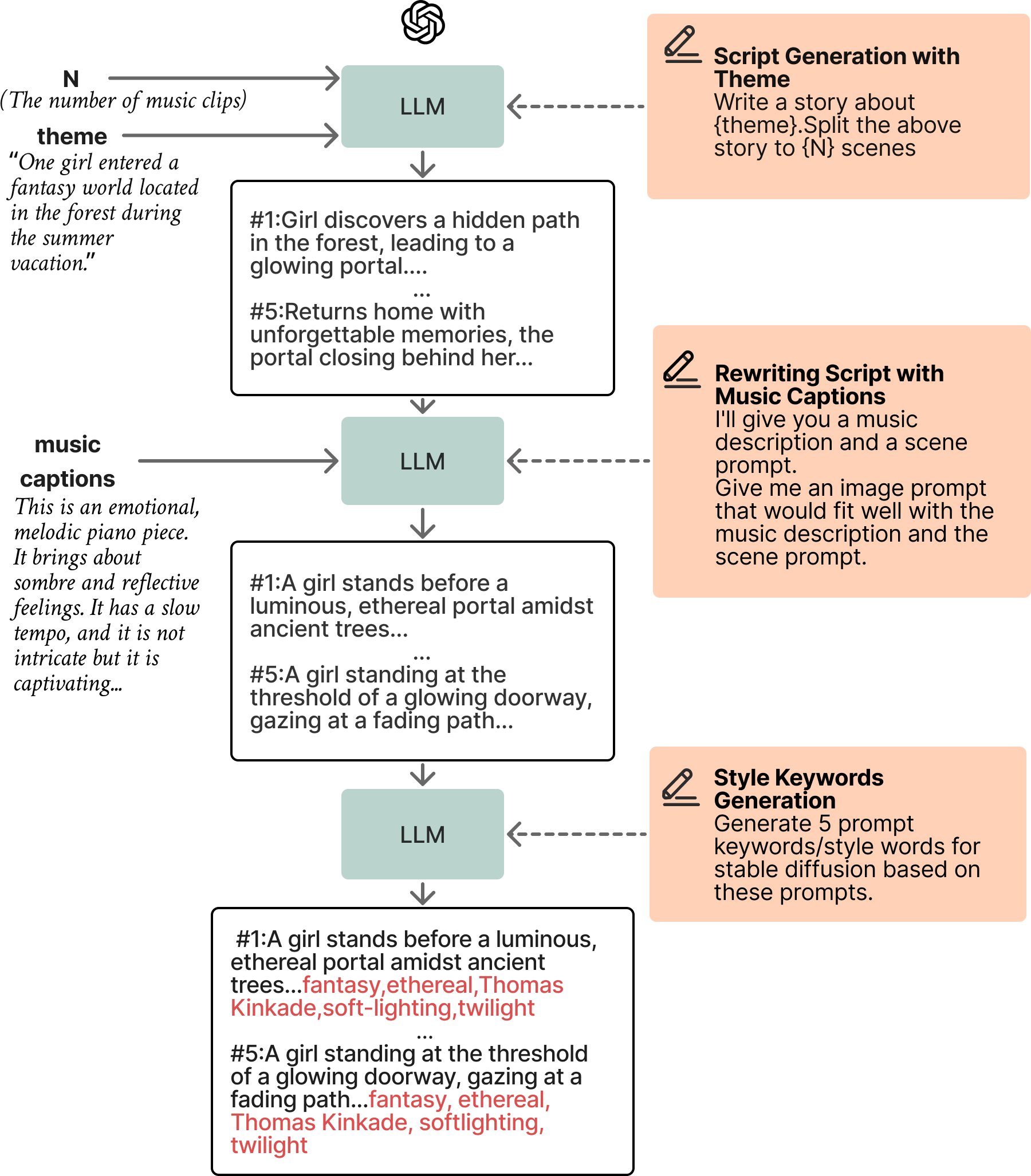}
    \caption{An example of three-step script generation with instructions of LLM. The input music is \textit{One Summer's Day} by Joe Hisaishi.}
    \label{fig:script-pipeline}
\end{figure}

\textbf{Case Study.} In Fig.~\ref{fig:script-pipeline}, we present an example of a three-step script generation process along with instructions for the LLM. Beginning with an input theme such as \textit{"One girl entered a fantasy world located in the forest during the summer vacation"} and accompanied by the input music \textit{One Summer's Day} by Joe Hisaishi, we proceed as follows. First, we guide the LLM to generate a descriptive story based on the theme, dividing it into $N$ scenes. Second, we feed each scene prompt and its corresponding music caption into the LLM, guiding it to generate an image prompt that fits the input information. Finally, the LLM is instructed to generate five style keywords based on all scene prompts. From the results, it is evident that the generated scenes and keywords vividly capture the serene ambiance of the input music. The complete script is available in Fig.~\ref{fig:script-example} within the appendix.

\subsection{Video Generation}
In this module, our task involves generating video clips for each scene in the script. Video generation models are mainly divided into two categories: text-to-video and image-to-video. If the text-to-video model is used, users must wait for the entire video generation process before editing unsatisfactory scenes. Conversely, by utilizing a text-to-image model to generate images, followed by an image-to-video model to generate video clips, users can preview the generated images and make edits accordingly. Given the significantly higher time cost of video generation, we opt for the second approach to enable users to preview visual content early, thus facilitating the editing process. Specifically, we employ an advanced text-to-image model Stable Diffusion XL 1.0~\cite{stablediffusion}, to transform the $N$ scenes in the script into $N$ images. Then, we utilize the state-of-the-art image-to-video model Stable Video Diffusion~\cite{stablevideodiffusion} to generate $N$ video clips from these images.

\subsection{Dynamic Music-Video Synchronization} 
\label{synchronization}

\begin{figure}[htp]
    \centering
    \setlength{\abovecaptionskip}{0.cm}
    \includegraphics[width=0.65\textwidth]{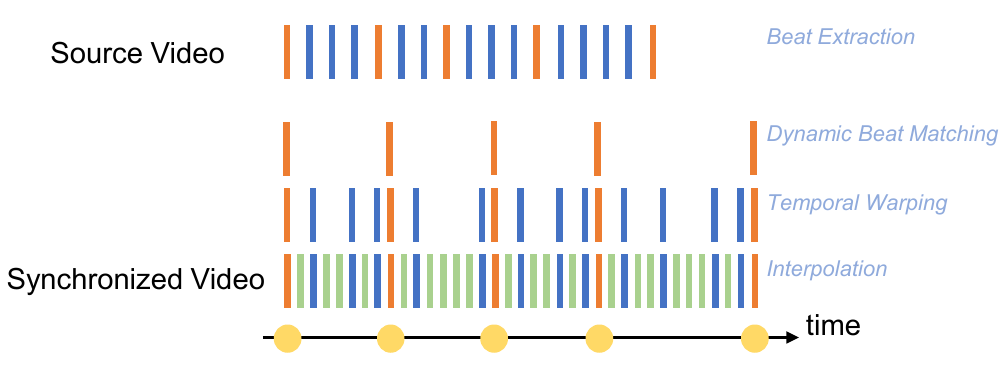}
    \caption{Overview of Dynamic Music-Video Synchronization. \textmd{
    \textcolor[RGB]{255,140,0}{Orange} indicates the key beat frames. \textcolor[RGB]{65,105,225}{Blue} indicates the remaining frames. \textcolor[RGB]{90,160,90}{Green} indicates the interpolated frames. \textcolor[RGB]{230,200,30}{Yellow} marks the music beats.}}
    \label{fig:pipeline-sync}
\end{figure}

After the video generation module processes the generated script, we obtain $N$ video clips. The next step is to rhythmically synchronize these clips with music clips and combine them into a music video. \highlight{This process relies on \textbf{\textit{visual beats}} - key visual events in the video that align with the music beats, such as scene transitions or character movements.} Due to the lower frame rate and frame count of AI-generated videos, the number and clarity of extracted visual beats are limited. When the number of music beats is higher, previous methods~\cite{Berndt1994UsingDT,davis2018visual} adopt a one-to-many pairing strategy, where a single visual beat corresponds to multiple music beats, leading to repeated video frames in synchronized videos. To address this issue, we propose a music-video synchronization method that avoids repeated frames. 
% In particular, we introduce a rule-based \textbf{visual rhythm extraction} method, followed by a \textbf{dynamic beat matching} algorithm aligning visual beats with music beats. Further, we devise a \textbf{temporal warping} strategy to enhance alignment and employ an \textbf{interpolation} technique for smoother synchronized video. 
\highlight{In particular, we employ a rule-based \textbf{visual rhythm extraction} method that identifies visual beats and visual impact envelopes. Building on this, we introduce a \textbf{dynamic beat matching} algorithm that dynamically aligns visual beats with music beats while maintaining temporal monotonicity, thus ensuring a more natural and synchronized flow between audio and video. Further, we devise a novel \textbf{temporal warping} strategy that aligns visual beats with music beats while enhancing visual effects around the beats. We also employ an \textbf{interpolation} technique to achieve smoother transitions in the synchronized video.}

% In scenarios with more music beats, previous methods~\cite{Berndt1994UsingDT,davis2018visual} adopt a many-to-one pairing strategy, resulting in repeated frames in synchronized videos. Therefore, we propose a music-video synchronization method that avoids repeated frames in the videos. We introduce a rule-based \textbf{visual rhythm extraction} method, followed by a \textbf{dynamic beat matching} algorithm aligning visual beats with music beats. Further, we devise a \textbf{temporal warping} strategy to enhance alignment and employ an \textbf{interpolation} technique for smoother synchronized video.

Our key idea is defining a \textbf{warping function} $\mathcal{G}: t \mapsto \mathcal{G}(t)$, which designates the timing $\mathcal{G}(t)$ for the frame from the source video at time $t$ to appear in the synchronized video. As shown in Fig.~\ref{fig:pipeline-sync}, the process begins with detecting and matching of key visual beat frames with the music beats. Subsequently, the remaining frames are warped to target times conditioned on the visual envelope. Finally, interpolation is applied to enhance the video's smoothness.

\textbf{Visual Rhythm Extraction.\label{sec:beat-extraction}}
% We extract visual rhythm features by capturing sudden visible deceleration as described in VisBeat~\cite{davis2018visual}.
\highlight{\textbf{\textit{Visual rhythm}} refers to the temporal arrangement of visible motion in a video, similar to how musical rhythm organizes sounds over time.} We utilize the visual rhythm extraction method described in VisBeat~\cite{davis2018visual}, which employs visual analogues of musical rhythm to characterize visual impact by capturing sudden visible deceleration. Specifically, we first calculate the directogram, represented by a 2-D matrix $D(i, \theta)$, to depict changes in visual magnitude across different angles.:
\begin{equation}
    D(i, \theta)=\sum\limits_{x,y}\vert F_i(x,y) \vert\mathbbm{1}_{\theta}(\angle F_i(x,y)),
\end{equation}
\begin{equation}
    \mathbbm{1}_{\theta}(\phi):=\left\{
    \begin{aligned}
    1 & \quad if \quad\vert \theta-\phi \vert \leq \frac{2\pi}{N_{bins}}  \\
    0 & \quad {\rm otherwise.}\\
    \end{aligned}
    \right.
\end{equation}
Here, $F_i(x,y)$ represents the optical flow field of the $i$-th frame computed using the Farneb{\"a}ck algorithm~\cite{farneback2003two} and $N_{\text{bins}}$ signifies the number of angular bins, set to 12 in our method.

\highlight{Next, we compute the \textbf{\textit{visual impact envelope}}, which captures the timing and intensity of significant visual changes in the video. This method is analogous to the \textbf{\textit{onset envelope}} in music, which measures the likelihood of an onset (a significant change) at each point in time. Onset envelopes in music typically correspond to an increase in spectral flux, marking the arrival of a new musical event or beat. Similarly, the visual impact envelope reflects key visual events in the video.} To compute it, we calculate the bin-wise deceleration $D_F(i,k)$, and sum over positive values in the columns of $D_F$ to get visual impact envelope $u^v(i)$ with the same form as the onset envelope of music:

% Next, we compute the bin-wise deceleration $D_F(i,k)$, and sum over positive values in the columns of $D_F$ to get visual impact envelope $u^v(i)$ with the same form as the onset envelope of music. Onset envelopes are a measure of how likely an onset, or significant change, occurs at each point in time. In music, they typically correspond to an increase in spectral flux, marking the arrival of a new musical event or beat. Similarly, the visual impact envelope captures the timing and intensity of key visual changes, reflecting the onset of significant visual events in the video, such as scene transitions or movements.
\begin{equation}
    D_F(i,k)=D(i,k)-D(i-1,k),
    %,\quad i =1,\cdots,n
\end{equation}
\begin{equation}
    u^v(i)=\sum\limits_{k=-\frac{N}{2}}^{\frac{N}{2}-1}\frac{D_F(i,k)+\vert D_F(i,k) \vert}{2}.
    %,\quad i =1,\cdots,n
\end{equation}

To detect discrete visual beats, we employ a 0.125-second window to identify local maxima in the impact envelope, selecting them as visual beats $\{v_n\}$ if they exceed their local mean by at least $10\%$ of the global maximum of the envelope. Due to the low frame rate (fps = 8) and frame count (25 frames) of AI-generated videos, making it challenging to extract tempograms, we omit the beat tracking method proposed in VisBeat~\cite{davis2018visual}.

\textbf{Dynamic Beat Matching.}
After extracting the visual impact envelope $u^v$ and visual beats $\{v_n\}$, we use \texttt{librosa}~\cite{mcfee2015librosa} to extract the music onset envelope $u^m$ and music beats $\{m_n\}$. In this section, we employ dynamic programming to find the optimal matching between the visual beats and the music beats. Assuming fewer visual beats than music beats ${\left|\{v_n\}\right| < \left|\{m_n\}\right|}$, our objective is establish one-to-one pairings between all visual beats and the music beats, allowing for unpaired music beats, while ensuring that the matching progresses monotonically forward in time. Under these constraints, we derive the cumulative distances $\gamma(i,j)$:

% Upon extracting the visual beats $\{v_n\}$, we match them with music beats $\{m_n\}$ using dynamic programming. The music onset envelope $u^m$ and music beats $\{m_n\}$ are obtained through \texttt{librosa}~\cite{mcfee2015librosa}.  
% Assuming fewer visual beats than music beats ${\left|\{v_n\}\right| < \left|\{m_n\}\right|}$, we aim to establish one-to-one pairings between all visual beats and music beats, allowing for unpaired music beats. Under these constraints, we derive the following cumulative distances $\gamma(i,j)$:

\begin{equation}
    \gamma(i,j)=d(i,j)+\underset{0\leq j'< j}{\mathrm{min}}\ \gamma(i-1,j'),
\end{equation}

\begin{equation}
    d(i,j) = \left|\frac{u^m(m_j)}{\overline{u^m}({m_n})} - \frac{u^v(v_i)}{\overline{u^v}({{v_n}})}\right| + \left|\frac{t_{m_j}}{T_m} - \frac{t_{v_i}}{T_v}\right|,
\end{equation}
Here, $t_{v_i}$ and $t_{m_j}$ denote the time of the $i$-th visual beat $v_i$ and the $j$-th music beat $m_j$. ${T_v}$ and $T_m$ are the duration of the video clip and the music clip. $\overline{u^v}({{v_n}})$ and $\overline{u^m}({m_n})$ indicate the average envelope of visual beats and music beats, respectively. $u^v$ and $u^m$ have been normalized to $[0,1]$. When there are fewer music beats, we interchange the positions of the music beat sequence and the visual beat sequence in the equations.

Upon solving the dynamic programming, we obtain the optimal matching beat pairs $\{(v'_i, m'_i)\}$, from which we deduce the warping of beat frames:
\begin{equation}
     \mathcal{G}(t_{v'_i}) = t_{m'_i}\label{equ:t},
\end{equation}
where $t_{m'_i}$ is the target time for source time $t_{v'_i}$.

\textbf{Temporal Warping.}
When the warping of beat frames is determined, we calculate the target times of the remaining frames based on the visual impact envelope $u^v$. Specifically, we first use linear interpolation to estimate the visual envelope at the intermediate moment $t$ between the $i$-th frame and $j$-th frame:
\begin{equation}
  u_{i,j}(t)= u^v{(i)}\frac{t-t_j}{t_i-t_j}+u^v{(j)}\frac{t-t_i}{t_j-t_i},
\end{equation}

Next, we replace the source time with the target time to estimate the visual envelope at $t$ in the warped video as
\begin{equation}
  u^{warp}_{i,j}(t)= u^v{(i)}\frac{t-\mathcal{G}(t_j)}{\mathcal{G}(t_i)-\mathcal{G}(t_j)}+u^v{(j)}\frac{t-\mathcal{G}(t_i)}{\mathcal{G}(t_j)-\mathcal{G}(t_i)}.
\end{equation}

Now, given $k$-th frame $(v'_i\leq k<v'_{i+1})$, we define $I_k$ as the envelope impulse between adjacent frames in Equation~\ref{eq:Ik}:
\begin{equation}
\label{eq:Ik}
\begin{aligned}
    I_k&=\int_{\mathcal{G}(t_k)}^{\mathcal{G}(t_{k+1})}u^{warp}_{k,k+1}(t) \text{d} t \\&=\frac{(\mathcal{G}(t_{k+1})-\mathcal{G}(t_k))(u^v{(k)}+u^v{(k+1)})}{2}.
\end{aligned}
\end{equation}

% To emphasize the synchronization of beat frames, we aim for smooth transitions between beat frames, which means that the impulse of each frame to be stable. Therefore, we set an objective function:

To emphasize the synchronization of beats, smooth transitions between beat frames are required, which means that the impulse between beat frames should remain stable. Therefore, we set the following objective function:

\begin{equation}
\min_{v'_i\leq k<v'_{i+1}} \textbf{Var} (I_k),
\label{equ:var_min}
\end{equation}
where \textbf{Var} represents the variance. 

% For any $k$ if $I_k=I_{k+1}$, Equation~(\ref{equ:var_min}) will get minimum. At this point, we can calculate $\mathcal{G}(t_k)$ by solving a linear system: 

For any $k$ if $I_k=I_{k+1}$, Equation~(\ref{equ:var_min}) will get minimum. In this case, we can derive a linear system from $I_k=I_{k+1}$ and compute the target time $\mathcal{G}(t_k)$ by solving this linear system:
\begin{align}
    &\mathcal{G}(t_k)=\mathcal{G}(t_{k+1})\alpha_k + \mathcal{G}(t_{k+2})\beta_k, \qquad k=v'_i,v'_i+1,\cdots, v'_{i+1}-2\label{eq:G_k},\\
    & \alpha_k=\frac{u^v(k)+2u^v(k+1)+u^v(k+2)}{u^v(k)+u^v(k+1)}, \\
    & \beta_k=-\frac{u^v(k+1)+u^v(k+2)}{u^v(k)+u^v(k+1)}.
\end{align}

% We can substitute $\mathcal{G}(t_{v'_i})$ and $\mathcal{G}(t_{v'_{i+1}})$ in Equation~\ref{eq:G_k} to obtain the target time for intermediate frames, thus achieving the warping of all video frames in the source video.  

By substituting the known target times of adjacent beat frames, $\mathcal{G}(t_{v'_i})$ and $\mathcal{G}(t_{v'_{i+1}})$, into the Equation~\ref{eq:G_k}, we can compute the target times for intermediate frames, thus achieving the warping of all video frames in the source video.

\begin{figure}[t]
    \centering
    \setlength{\abovecaptionskip}{0.cm}
\includegraphics[width=0.7\textwidth]{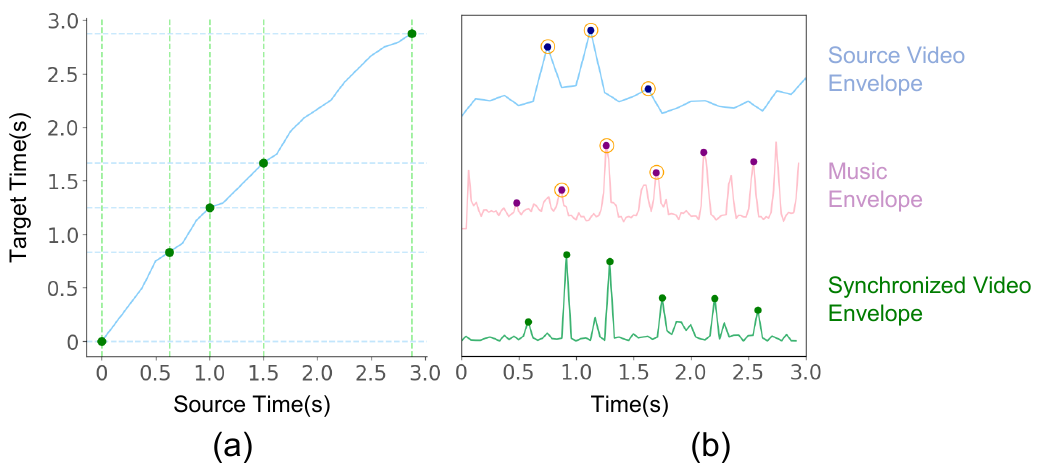}
    \caption{Visualization of synchronization outcomes: (a)\textmd{ The \textcolor[RGB]{100,150,240}{warping curve} illustrates the mapping from source \( t \) to target \( \mathcal{G}(t) \), with \textcolor[RGB]{0,120,0}{green} dots representing the matching of visual beats with music beats.} (b)\textmd{ Comparative envelope curves for the \textcolor[RGB]{100,150,240}{source video}, the \textcolor[RGB]{250,150,150}{music}, and the \textcolor[RGB]{40,180,120}{synchronized video}. Detected beats are marked by dots (\(\bullet\)) while the matched beat pairs are encircled (\textcolor[RGB]{255,140,0}{\(\bigcirc\)}).}}
    \label{fig:sync-result}
    % \vspace{-0.5cm}
\end{figure}

\textbf{Interpolation.}
After warping source video frames to target times, there are remaining unfilled frames in the synchronized video. Thus, we use an interpolation model, RIFE~\cite{huang2022rife}, to generate intermediate frames, filling these gaps and increasing the frame rate to 24 for smoother playback. With interpolation completed, we achieve the generation of a synchronized video. Fig.~\ref{fig:sync-result} is an example result produced by our method, showcasing well-aligned visual beats with music beats after synchronization.

Once all music clips and video clips are synchronized, we concatenate the synchronized videos together, ultimately producing a multi-scene music video.

\subsection{Implementation Details}
We implement the MV-Crafter full-stack system using Python Flask and Javascript React, deploying it on a server equipped with an NVIDIA A800 GPU. In the script generation module, we apply GPT-4~\cite{openai2023gpt4} to create music video scripts, leveraging the music captions generated by LP-MusicCaps~\cite{Doh_Choi_Lee_Nam_2023}. In the video generation module, we first utilize Stable Diffusion XL 1.0~\cite{stablediffusion} as a text-to-image generator to generate scenes at a resolution 576x1024. These scenes are then fed into the Stable Video Diffusion (SVD) Image-to-Video model~\cite{stablevideodiffusion} to produce 25 frames of videos at fps 8, with the $motion\_bucket\_id$ increased to 200 for pronounced motions. For the music-video synchronization module, we utilize \texttt{librosa}~\cite{mcfee2015librosa} to extract music features and calculate visual rhythm features based on optical flow extracted using OpenCV~\cite{farneback2003two}. We further interpolate frames to warp the video and change fps from 8 to 24 by an advanced interpolation model RIFE~\cite{huang2022rife}.

\section{Experiments}
We first conducted a technical evaluation to assess the synchronization quality and visual quality of the generated music videos. To further validate the effectiveness of the synchronization module, we compared it with other synchronization algorithms by evaluating the degree of beat matching and visual effects.

\subsection{Setup}
\textbf{Data.} We collected 20 rhythmic music pieces with different tempos from Spotify and YouTube, ranging from 69 to 166 bpm (beats per minute). The average duration of the music pieces was 80.5 seconds. The selected pieces fell into ten genres, including pop, electronic, country, classical, dance, soul, hip-hop, classical, Christmas, and funk. We crafted themes with an average length of 14 words based on various musical cues including emotions, genres, and cultural contexts, resulting in 20 music-theme pairs.

\textbf{Baselines.} Due to the absence of a similar pipeline that can automatically generate music videos from input themes and music, we constructed two baselines based on existing frameworks: Generative Disco + LLM and Music2Video + LLM. Generative Disco~\cite{Liu2023GenerativeDT} is a generative music visualizer that utilizes GPT-4~\cite{openai2023gpt4, wu2022wav2clip} and Stable Diffusion Videos~\cite{stablediffusionvideo}, requiring users to select the music intervals and input each interval's visual description. Music2Video~\cite{Jang2022Music2VideoAG} employs CLIP models~\cite{radford2021learning, wu2022wav2clip} to convert input music and lyrics into latent, guiding VQ-GAN~\cite{esser2021taming} in generating music videos. Since these frameworks lack music segmentation and script-writing capabilities, we divided intervals according to music beats extracted by \texttt{librosa}~\cite{mcfee2015librosa} and utilized GPT-4 to generate scripts solely based on the theme instead of our three-step script generation module. The scripts were fed into Generative Disco and Music2Video, replacing the visual descriptions and lyrics, respectively. 

\textbf{Metrics.} We evaluated our method from two perspectives. First, we introduced the Beat Alignment Score~\cite{Li2021AICM} to assess the degree of synchronization between the music and video. Second, we validated the correspondence between video content and input theme using the CLIPSIM~\cite{Wu2021GODIVAGO} metric.

\textit{Beat Alignment Score.} Beat Alignment Score (BAS)~\cite{Li2021AICM} evaluates motion-music alignment by measuring the similarity between motion beats and music beats in dance generation. In our experiments, we replace motion beats with extracted visual beats and compute the music beats by \texttt{librosa}~\cite{mcfee2015librosa}. BAS is defined as the average distance between each visual beat and its nearest music beat:
\begin{equation}
    BAS=\frac{1}{\lvert{B^v}\rvert}\sum_{i=1}^{\lvert{B^v}\rvert}{\rm exp}\left(-\frac{\mathop{\min}_{{\forall}t_{j}^m{\in}B^m}{\Vert t_i^v-t_j^m \Vert^2}}{2\sigma^2}\right),
\end{equation}
where $B^v=\{t_i^v\}$ represent visual beats, $B^m=\{t_j^m\}$ represent music beats and $\sigma$ is set to $0.1$ to normalize sequences.

\textit{CLIPSIM.} CLIPSIM~\cite{Wu2021GODIVAGO} evaluates the similarity between the generated video and the input text. We apply the pre-trained CLIP~\cite{radford2021learning} ViT-H/14 model to calculate the similarities between theme and video frames and then take the average value as below:

\begin{equation}
    SIM(t,v)=\frac{1}{L}\sum\limits_{l=1}^{L}CLIP(t,v^{(l)}),
\end{equation}
where $t$ denotes the input theme, $v^{(l)}$ is the $l$-th frame in the generated music video.

\begin{figure*}
  \centering
  \includegraphics[width=\textwidth]{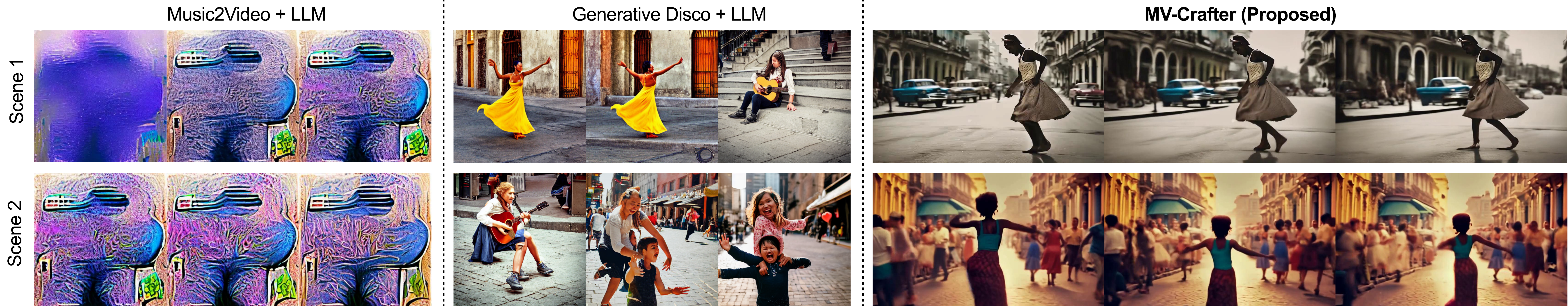}
  \caption{Qualitative comparisons with baselines. We show the results of two scenes in the generated music videos for the song \textit{Havana} by Camila Cabello.}
% \vspace{-0.3cm}
  \label{fig:qualitative}
\end{figure*}

\begin{figure}[t]
  \centering
  % \vspace{-0.0cm}
  \includegraphics[width=0.65\textwidth]{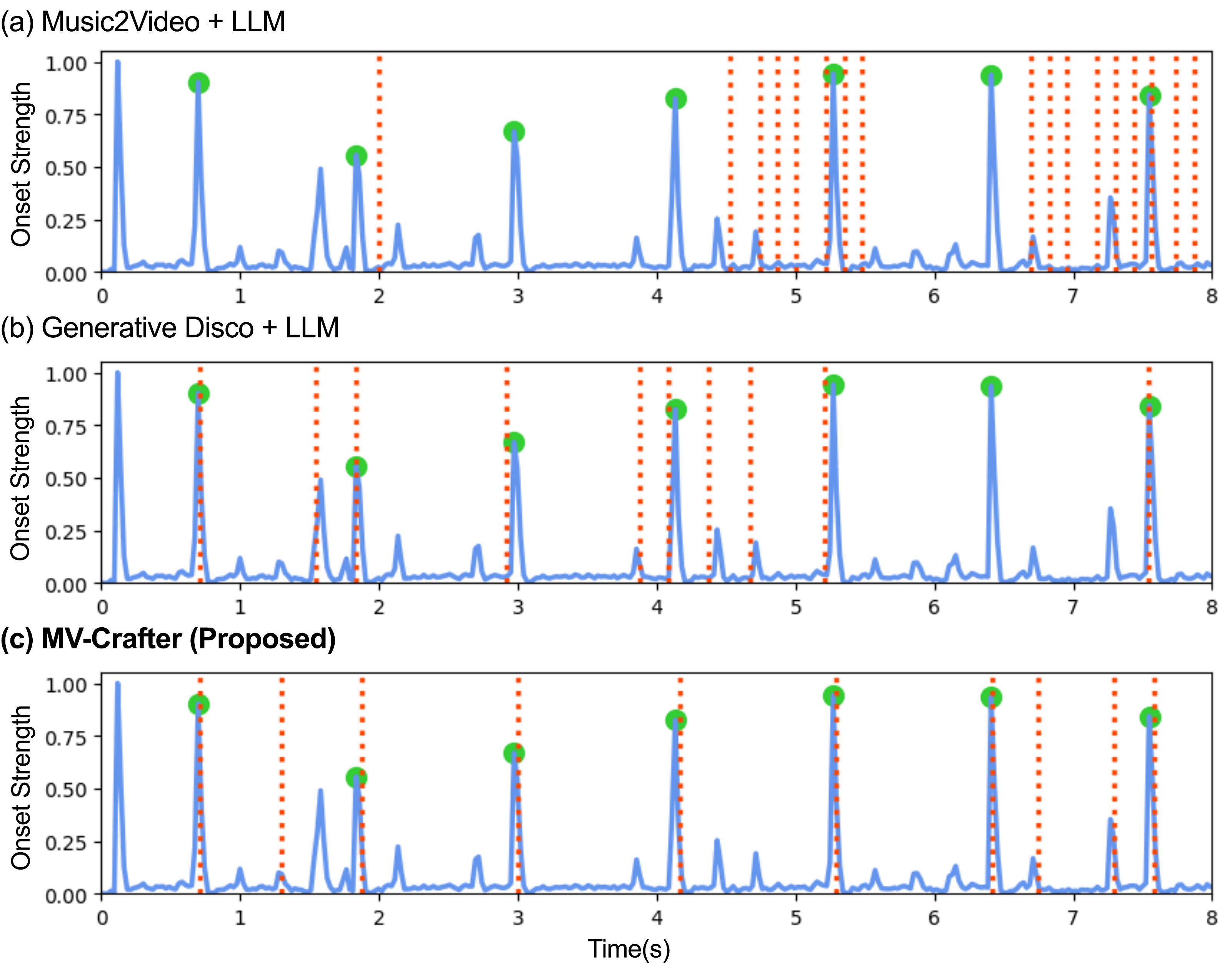}
  \caption{Qualitative comparisons on synchronization. The \textcolor[RGB]{100,149,237}{blue} curve is the music onset envelope, \textcolor[RGB]{50,205,50}{green} dots represent the music beats. The \textcolor[RGB]{255,69,0}{red} dotted lines indicate the visual beats of the synchronized video. Aligning more visual beats with music beats indicates a better synchronization effect.}
  % \vspace{-0.5cm}
  \label{fig:qualitative_curve}
\end{figure}

\subsection{\highlight{Qualitative Example: A Woman is Dancing in Havana}}
We visualized generation results for the music \textit{Havana} and theme \textit{"A woman is dancing in Havana"} in Fig.~\ref{fig:qualitative} for comparison. It was evident that Music2Video, employing VQ-GAN, struggled to interpret text embeddings from complex scripts, resulting in meaningless visual content. Generative Disco generated frames individually through latent walk in diffusion space, posing challenges in producing coherent shots. In contrast, our approach combined the text-to-image model with the image-to-video model, effectively generating high-quality visuals and coherent motions in video scenes.

Fig.~\ref{fig:qualitative_curve} illustrates the alignment between music beats and visual beats in synchronized music videos. Our method accurately aligned visual beats with all music beats, confirming the effectiveness of the music-video synchronization module. Through visual envelope-induced warping, we ensured smooth transitions between key beat frames, achieving better synchronization with the rhythm points. 

% \begin{table}[t]
% % \vspace{-0.0cm}
%   \caption{Quantitative comparisons with competing methods.}
%   \label{tab:quantitative}
%   \small
%   \begin{tabular}{cr@{}lr@{}l}
%     \toprule
%     Method & B&AS $\uparrow$ & CL&IPSIM $\uparrow$\\
%     \midrule
%     Music2Video + LLM & 0&.4917 & 8&.7434\\ 
%     Generative Disco + LLM & 0&.6224  & 15&.6679\\ 
%     \textbf{MV-Crafter (Ours)} & \textbf{0}&.\textbf{7605} & \textbf{18}&.\textbf{2679}\\ 
%     \bottomrule
%   \end{tabular}
%   % \vspace{-0.3cm}
% \end{table}

\begin{figure}[t]
% \vspace{-0.3cm}
  \centering
  \includegraphics[width=1\textwidth]{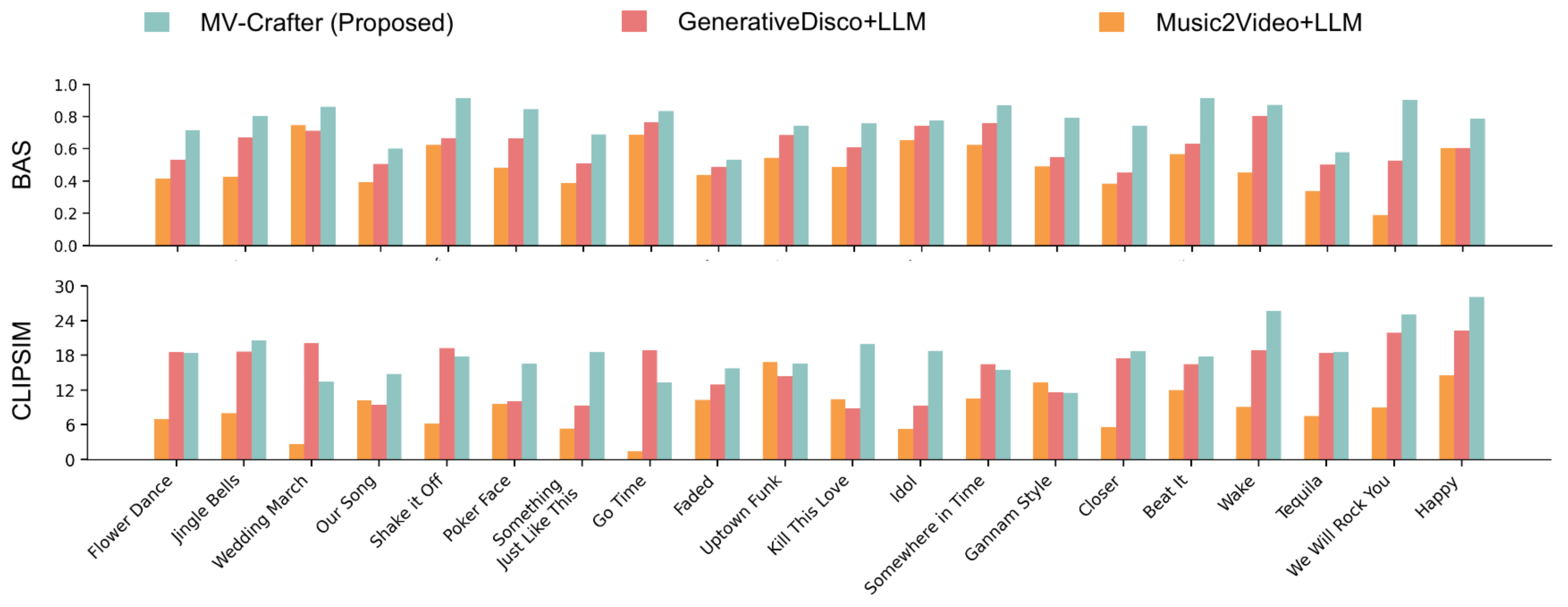}
  \caption{\highlight{Quantitative comparisons with competing methods based on BAS and CLIPSIM scores. Each bar represents the metric computed for a music video generated from a specific music-theme pair using one of the three methods.}}
  \label{fig:quantitative}
  % \vspace{-0.5cm}
\end{figure}

\subsection{Quantitative Results}
\highlight{We took 20 prepared music-theme pairs as inputs and generated 60 music videos using three methods. Subsequently, we calculated the BAS and CLIPSIM metrics for the generated results. As shown in Fig.~\ref{fig:quantitative}, each music video is represented by three bars, corresponding to the metrics computed for the three methods.}

\highlight{Focusing on the BAS metric, our proposed method outperformed the other two methods across all music videos. Specifically, the average BAS for MV-Crafter was significantly higher $(M = 0.777, SD = 0.111)$, compared to GenerativeDisco+LLM $(M = 0.618, SD = 0.106)$ and Music2Video+LLM $(M = 0.496, SD = 0.135)$, which demonstrates the consistent effectiveness of the proposed synchronization method.}

\highlight{Next, we turn to the CLIPSIM metric. MV-Crafter outperformed the other methods, with 13 out of the 20 videos achieving the highest score. The average CLIPSIM score for MV-Crafter $(M = 18.268, SD = 4.172)$ was higher than that of GenerativeDisco+LLM $(M = 15.668, SD = 4.514)$ and Music2Video+LLM $(M = 8.743, SD = 3.835)$, further highlighting its ability to generate content closely aligned with the input theme.}

\highlight{Despite the overall strong performance,we identified certain videos, such as \textit{Wedding March}, that scored lower in CLIPSIM. Upon reviewing the generated video, we found that the MV-Crafter version contained fewer lawn scenes and more ornate, classical banquet hall settings. This discrepancy may stem from our three-step script generation method, which integrates music captions. By incorporating the musical style into the script generation process, the method might modify the scene slightly, prioritizing visual elements that align with the music’s aesthetic rather than strictly adhering to the input theme. Future improvements could focus on refining the balance between adhering strictly to the input theme and integrating musical style, ensuring both thematic relevance and aesthetic harmony in the final video.}

% As presented in Fig.~\ref{fig:quantitative}, our method surpassed the competing methods in both metrics, demonstrating the effectiveness in generating thematically relevant and rhythmically engaging music videos.

% \begin{table*}[]
%     \caption{User study on rhythm synchronization, video-theme correspondence, video-music correspondence, video content coherence, comprehensive quality.}
%     \small
%   \begin{tabular}{cccccl}
%     \toprule
%     Method & Synchronization & Theme Correspondence & Music Correspondence & Coherence & Quality \\
%     \midrule
%     Music2Video + LLM & 1.372 & 1.114 & 1.226 & 1.428 & 1.154 \\
%     Generative Disco + LLM & \textbf{3.414} & 3.120 & 2.992 & 2.512 & 2.732 \\
%     SVD + LLM & 2.940 & 2.926 & 2.818 & 2.388 & 2.624 \\
%     \textbf{MV-Crafter (Ours)} & 3.260 & \textbf{3.817} & \textbf{3.855} & \textbf{3.505} & \textbf{3.507} \\
%     \bottomrule
%   \end{tabular}
%     \label{tab:user-study}
%     % \vspace{-0.5cm}
% \end{table*}

\subsection{Comparisons of Synchronization Methods}
\begin{figure}[t]
% \vspace{-0.3cm}
  \centering
  \includegraphics[width=1.0\textwidth]{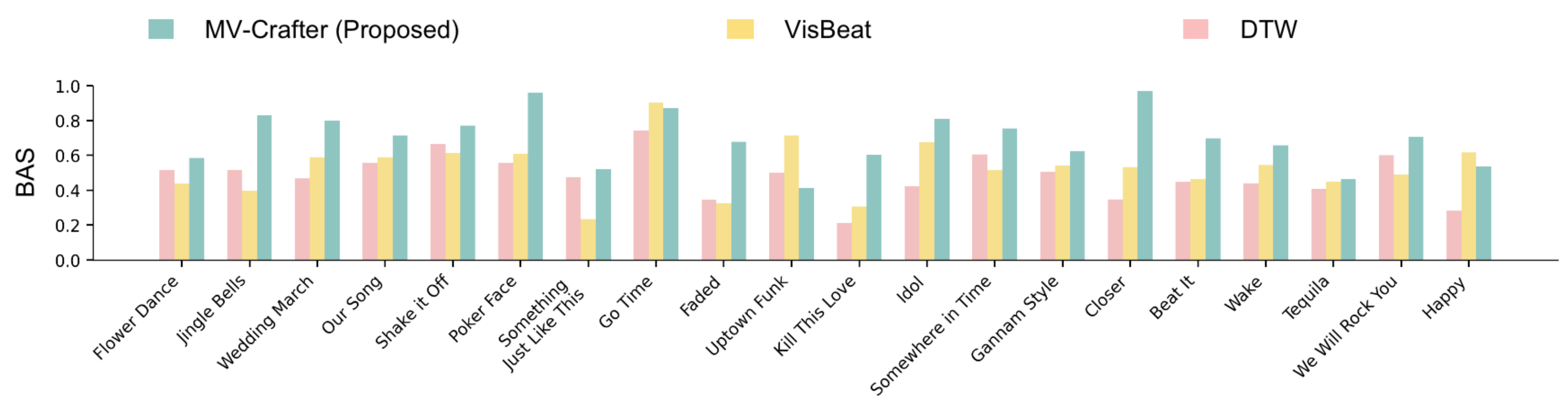}
  \vspace{-0.5cm}
  \caption{\highlight{Quantitative results of synchronization methods based on BAS. Each bar represents the BAS for a music video generated from a specific music-theme pair using one of the three synchronization methods.}}
  \label{fig:ablatio_bas}
  % \vspace{-0.5cm}
\end{figure}

To demonstrate the effectiveness of our synchronization module, we compared it with two classical approaches: VisBeat~\cite{davis2018visual} and dynamic time warping (DTW)~\cite{Berndt1994UsingDT}. VisBeat employs an unfolding strategy that generates random walks through the visual beats of the source video to align with target music of arbitrary length. The random walk's path is determined by a momentum parameter $\phi$, with a $0.5+\phi$ probability moving forward and a $0.5-\phi$ probability moving backward at each position. We used the default value of 0.1 in our experiment. DTW finds the optimal warping path to align time series with different lengths and rhythms. We segmented the prepared 20 music pieces into 372 music clips by our segmentation method and generated 372 video clips based on the generated scripts. \highlight{We then synchronized these video clips with their corresponding music clips using three synchronization methods and assessed the results by computing beat alignment score (BAS) for each music video, as shown in Fig.~\ref{fig:ablatio_bas}. Our method achieved the highest average BAS of 0.699 $(M = 0.699, SD = 0.153)$, outperforming VisBeat $(M = 0.528, SD = 0.152)$ and DTW $(M = 0.481, SD = 0.127)$. We further analyzed the video \textit{Uptown Funk} with a lower BAS. Upon review, we found that most music clips in this video were relatively long (over 5 seconds), while the generated video clips were only 3.125 seconds. Our synchronization method uses interpolation to ensure monotonous temporal alignment, but the increased interpolation may have reduced the visual prominence of the beat, leading to the lower BAS. However, this issue was observed in only a few  cases, and the significantly higher average BAS across all videos underscores the effectiveness of our synchronization method.}

% \begin{table}[t]
%     % \vspace{-0.2cm}
%   \caption{Quantitative results of synchronization methods}
%   \label{tab:ablation synchronization}
%   \small
%   \begin{tabular}{cccc}
%     \toprule
%     Method & VisBeat & DTW & \textbf{Ours}  \\
%     \midrule
%     BAS $\uparrow$ & 0.4835 & 0.4893 & \textbf{0.6642} \\
%     \bottomrule
%   \end{tabular}
%   % \vspace{-0.5cm}
% \end{table}

\begin{figure}[t]
% \vspace{-0.3cm}
  \centering
  \includegraphics[width=0.7\textwidth]{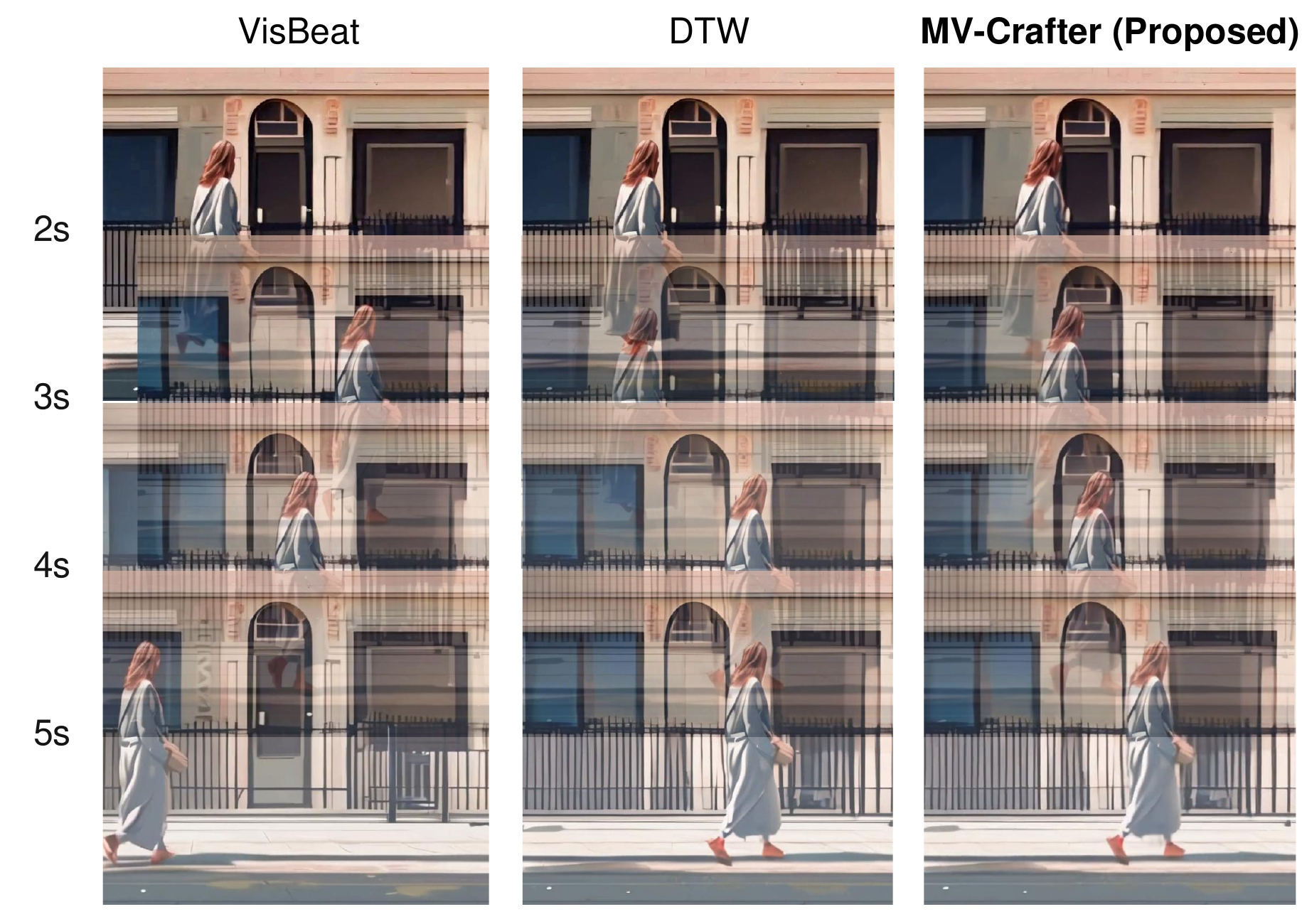}
  \caption{Comparison of synchronized videos on a 3.1-second original video clip with a 5.3-second music clip. Rather than matching with repetitive movements or keeping still, our method can align with music monotonically.
  % The original video clip depicts a scene where a woman is walking forward. We showcase frames from the synchronized videos at 2, 3, 4, and 5 seconds.
  }
  \label{fig:ablation-sync}
  % \vspace{-0.5cm}
\end{figure}

Fig.~\ref{fig:ablation-sync} illustrates the visual differences in synchronized videos. In the VisBeat result, the woman moved forward from 2 to 3 seconds but retreated to the original position from 3 to 5 seconds. This was because the unfolding strategy extended short videos by combining forward and backward steps on the timeline to match the longer music. Although effective for dancification tasks, we aimed to avoid repetitive actions in music videos. We also tried setting $\phi=0.5$ for Visbeat to progress forward monotonically in time. However, the synchronized video length was cut to 1.0s, indicating that Visbeat's monotonic progression couldn't perfectly align the video and music. In the DTW result, the video remained static at 2 to 3 seconds and 4 to 5 seconds due to the lack of path constraints. This led to one visual beat corresponding to multiple music beats, causing a paused effect in synchronized videos. In contrast, our dynamic beat matching method restricted one-to-one alignment, achieving beat synchronization without repetitive video frames. We then extended the short video through the temporal warping and interpolation process, resulting in a synchronized music video with temporal monotonicity, fluid visual scenes, and accurately matched beats.

\section{User Study}
% Despite the above quantitative evaluation of the proposed synchronization technique, we conducted two user studies to evaluate the quality of the generated music videos and the effectiveness of the key steps in generating music video scripts, which quantitative experiments can hardly estimate. 

\highlight{Despite the above quantitative evaluation of the proposed synchronization technique, we conducted four user studies to thoroughly evaluate the performance and user experience of the proposed system, which quantitative experiments can hardly estimate. These studies aimed to assess the quality of the generated music videos by comparing different generation methods, evaluate the effectiveness of key steps in generating music video scripts, and finally, investigate user experiences in system usage.}

%Specifically, we first evaluate the quality of the generated music video from five aspects that are closely relevant to our contributions. An ablation study was also performed to test the effectiveness of each of the three key steps in generating music scrept.

\subsection{Estimation on Music Video Quality}
In this study, we aimed to verify the quality of the generated music video based on the proposed technique by having participants rate based on five metrics through a questionnaire survey. 

\textbf{Baselines.} Three baselines were used in our experiment. Despite the aforementioned Generative Disco + LLM and Music2Video + LLM techniques, we utilized the SVD (Stable Video Diffusion) + LLM as the third baseline to eliminate the potential influence caused by using different video generation models. This approach directly concatenated the resulting video clips to generate a music video without synchronization to the music. In all these baselines, we applied the same method to generate music video scripts as introduced in Section 5.1. 
% In all these baselines, we use the same method to generate music video scripts as introduced in Section 5.1 and the result video clips are directly concatenated to generate a music video without synchronizing to the music. 

\textbf{Materials.}
We collected 6 rhythmic music tracks from Spotify, spanning 4 genres with tempos ranging from 69 to 166 bpm. For each music track, we designed corresponding themes based on musical cues, including emotions, genres, and cultural contexts, to ensure a diverse range of topics for the resulting music videos. Consequently, 24 music videos (6 music tracks $\times$ (3 baselines + 
1 MV-Crafter)) with an average length of 53 seconds were generated for comparison.

\textbf{Metrics.}
We introduce five metrics to evaluate the overall quality of the generated music videos according to the design requirements. To specifically evaluate our primary contributions, i.e., music video script generation and music-video synchronization, the following metrics that estimate the quality of the music video from different aspects were used in our experiment:
\begin{itemize}
  \item \underline{\textit{Synchronization}}: estimates how well the video rhythm synchronized with the music rhythm.
  \item \underline{\textit{Theme Correspondence}}: estimates how well is the video content correlated with the input theme.
  \item \underline{\textit{Music Correspondence}}: estimates how well the visual style (e.g., lighting, color, layout) correlated to the music style (e.g., genre, emotion, rhythm).
  \item \underline{\textit{Narrative}}: estimates how well are the succeeding scenes logically connected.
  \item \underline{\textit{Quality}}: estimates the overall quality of the music video.
\end{itemize}

\textbf{Participants.}
We recruited 30 participants (13 males and 17 females) aged 21 to 45 years old $(M=24.8, SD=4.5)$ for our study, including students, researchers, and engineers from diverse majors such as design, computer science, and electronic engineering. 14 ($86.7\%$) participants reported to watch music videos monthly or even weekly. Their expertise in creating music videos is as follows: novice: 17 ($56.7\%$), beginner: 9 ($30.0\%$), competent: 4 ($13.3\%$).

\textbf{Procedure.}
We conducted a within-subject study, in which a user needed to rate each generated music video through an online questionnaire. The generated music videos were divided into 6 groups based on the input music, with the order of videos within each group randomized. Participants were informed that all the videos were generated by AI and were asked to provide subjective ratings for five metrics on a 5-point Likert scale (1="very poor", 5="very good") after watching the music videos. 

\begin{figure*}
    \centering
    \includegraphics[width=1.0\textwidth]{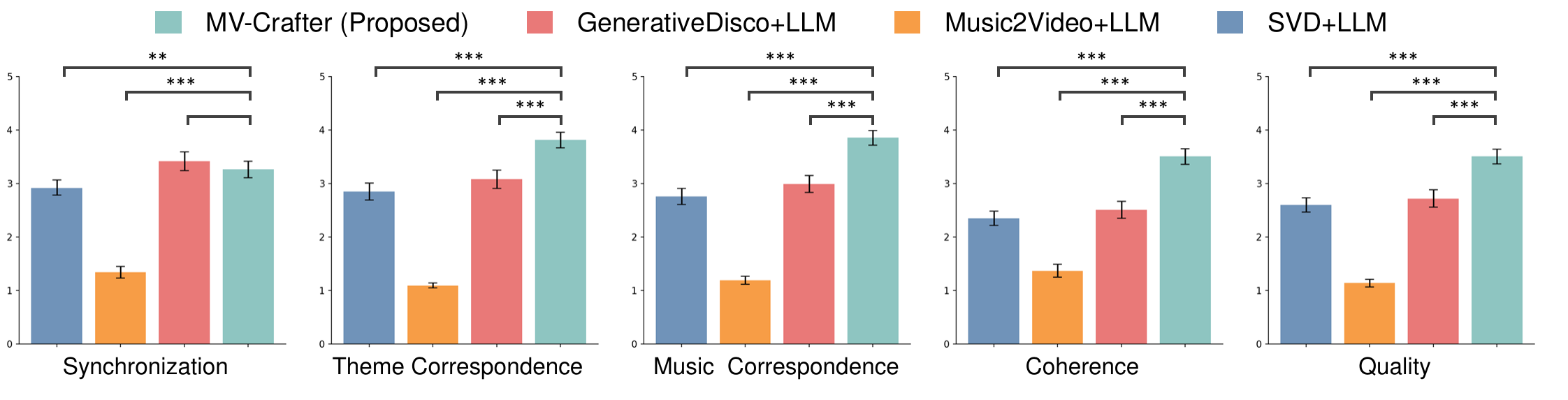}
    \caption{The ratings of music videos on rhythm synchronization, video-theme correspondence, video-music correspondence, video narrative, overall quality. Horizontal brackets indicate pairwise significant difference ($*: p<.05, **: p<.01, ***: p<.001$). The error bars represent the $95\%$ confidence intervals.}
    \label{fig:user-study}
\end{figure*}

\textbf{Results.} 
The results of the user study are presented in Fig.~\ref{fig:user-study}. We conducted a one-way ANOVA analysis that revealed significant differences among all the metrics ($p<.001$). Subsequently, we performed a post-hoc Tukey multiple comparison test to examine pairwise differences between MV-Crafter and other methods. The results indicated that our method significantly outperformed other approaches in terms of theme correspondence ($M=3.82, SD=0.97$), music correspondence ($M=3.82, SD=0.97$), narrative ($M=3.82, SD=0.97$), and the overall quality ($M=3.82, SD=0.97$) ($p<.001$ for all comparisons). Both MV-Crafter ($p<.01$) and Generative Disco+LLM ($p<.01$) are significantly better than SVD+LLM in the rating of synchronization.

%While Generative Disco+LLM ($M=3.42, SD=1.20$) slightly surpassed our method ($M=3.27, SD=1.06$) in the synchronization without a significant ($p=.495$ ). This can be attributed to the stable diffusion walk employed in Generative Disco, allowing each frame's generation to be controlled by the music. In contrast, our method achieves synchronization by warping the generated video to the music with a coarser synchronization granularity. For most music videos, achieving “cut-to-the-beat” and synchronization of music and visual beats already produces satisfactory visual results. Overemphasis on rhythm may lead to frequent changes in visual content, reducing video coherence and visual quality. 

%Compared to SVD+LLM, our method was rated significantly higher on all metrics ($p<.01$), demonstrating the effectiveness of the design of the synchronization module and script generation module.

\begin{figure}
    \centering
    \includegraphics[width=0.5\textwidth]{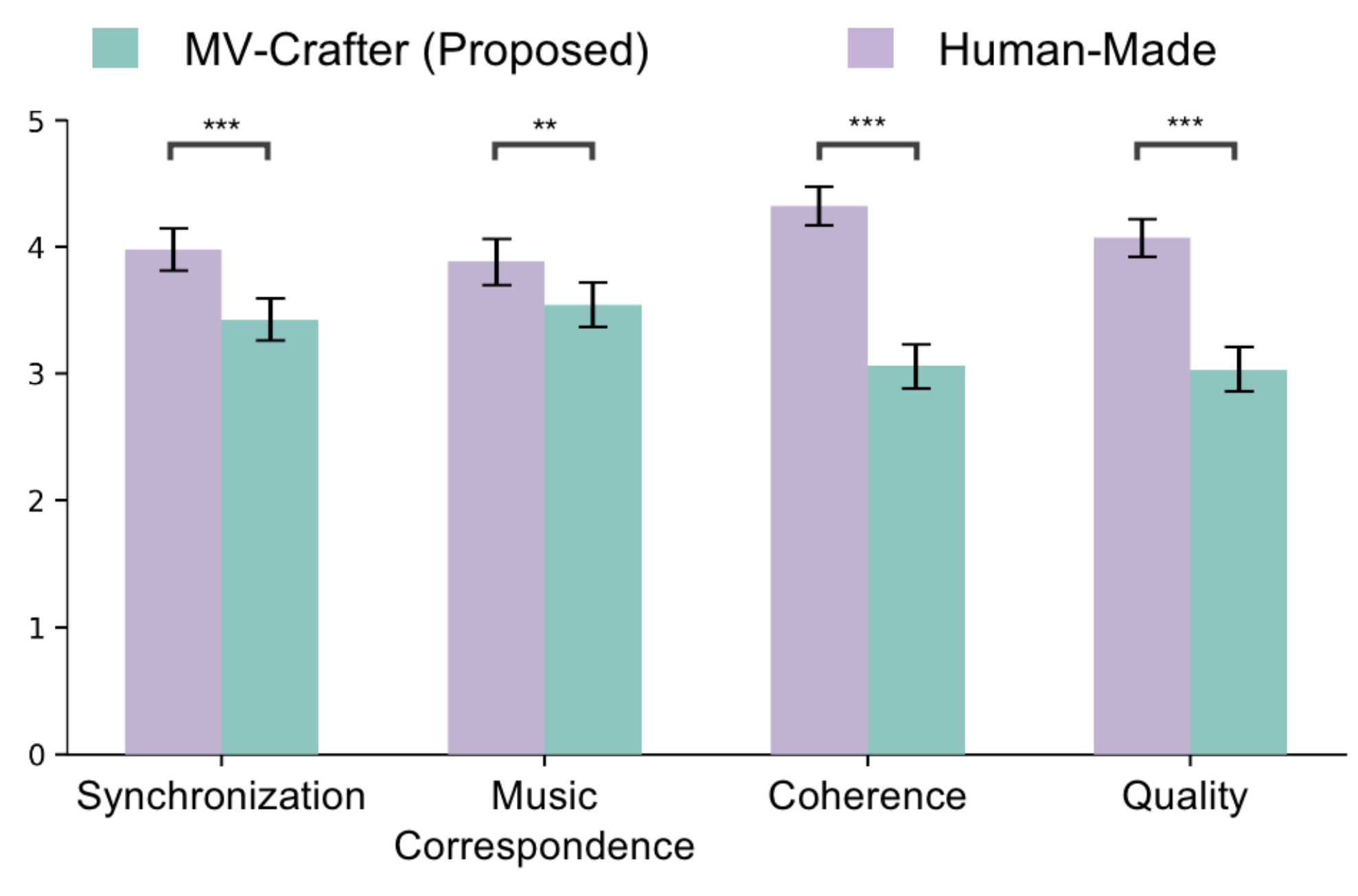}
    \caption{\highlight{The results of comparing MV-Crafter-generated and human-made music videos. Horizontal brackets indicate pairwise significant difference ($*: p<.05, **: p<.01, ***: p<.001$). The error bars represent the $95\%$ confidence intervals.}}
    \label{fig:human-ai}
\end{figure}

\subsection{\highlight{Comparison with Human-Made Music Videos}} 
\highlight{In this experiment, we extended our evaluation to compare the quality of the music videos generated by MV-Crafter system with that of human-made animated music videos.}

\highlight{\textbf{Materials.} Given the significant style differences between AI-generated scenes and real-life videos, we chose to compare the generated music videos with animated music videos instead of live-action ones. We collected five human-made animated music videos from YouTube, each with over 10 million views. Based on the content of these videos, we identified five themes and used MV-Crafter to generate music videos with corresponding themes and music. This resulted in five groups of music videos, each created by two different methods: AI-generated and human-made.}

\highlight{\textbf{Participants and Procedure.} We recruited 25 participants (15 females, 10 males) aged 18-31 years old ($M=23.6, SD=3.51$) and then conducted a within-subject study, similar to the procedure outlined in Section 6.1. Participants were not informed about the generation method of each music video and were asked to rate the music videos through a questionnaire. Since human-made music videos were not created with specific themes in mind, we excluded the \textit{theme correspondence} metric. The music videos were evaluated based on four metrics: \textit{synchronization, music correspondence, narrative} and \textit{quality}. To further understand users' perceptions of the two video types, participants were informed at the end of the questionnaire that each group contained both an AI-generated and a human-made music video, and their feedback was collected accordingly.}

\highlight{\textbf{Results.} The questionnaire results were analyzed using a one-way ANOVA with post-hoc Tukey tests. The analysis revealed that, while the MV-Crafter system produced videos that were reasonably aligned with the music, it consistently lagged behind human-made music videos across all evaluated metrics, as shown in Fig.~\ref{fig:human-ai}.}

\highlight{Further analysis of the user feedback revealed several key areas for improvement in the MV-Crafter system. Participants generally noted that while MV-Crafter-generated videos exhibited an impressive alignment with the music's rhythm and beats, they often lacked the narrative coherence and stylistic consistency that are characteristic of human-made videos. Despite maintaining a similar visual style, inconsistencies in environments and characters can make scenes appear disconnected and lacking logical flow. Additionally, several participants also highlighted technical issues with the MV-Crafter-generated videos, such as low-resolution visuals and awkward animation.} 

\highlight{It is important to recognize that human-made music videos benefit from extensive experience, creativity, and iterative refinement—factors that are challenging for AI systems to replicate at this stage. The current limitations in video generation technology, such as challenges in controlling stylistic consistency, character alignment, and avoiding motion blur, contribute to these discrepancies. Nevertheless, the MV-Crafter system demonstrates substantial potential, particularly in aligning video content with music rhythm, and with advancements in video generation technology, we expect significant improvements in both narrative coherence and technical quality.}

\subsection{Estimation on Script Generation}
We conducted an ablation study to evaluate the proposed music video script generation method as it directly affects the quality of video generation. 

% \textbf{Baselines.} We ablate the proposed script generation methods (denoted as $G$), producing two baselines. Specifically, the first baseline ($B1$) generates scripts without using style keywords, i.e. G (w/o style keywords). The second baseline $B_2$ generates scripts without using both style keywords and music captions, i.e., G (w/o script keywords and caption). 

\textbf{Baselines.} We ablate the proposed script generation method, producing two baselines. Specifically, the first baseline ($B1$) ablates the style keyword generation step of the script generation method, resulting in generated scripts without style keywords, i.e. w/o style keywords. The second baseline $B_2$  further ablates the step of rewriting the script using music captions, generating scripts based solely on the theme, i.e., w/o music captions, style keywords. 

\textbf{Materials.} Using the above baselines and the proposed script generation method, we generated 15 music video scripts based on 5 pieces of music in different genres that were downloaded from Spotify. To facilitate reading and comparison, we visualize each script by a list of scene images that were generated based on Stable Diffusion XL-1.0~\cite{stablediffusion} using the scene prompts in each script.

%Since the script content serves as descriptive prompts suitable for text-to-image generation, we utilized Stable Diffusion XL-1.0~\cite{stablediffusion} to convert all scripts into lists of images, facilitating evaluation by users based on the scenes.

%We selected five music pieces with different genres from Spotify. To prevent potential influence on script generation from the inherent correlation between the theme and the music, we devised themes unrelated to the musical content. Subsequently, we generated 15 scripts using three methods. Since the script content serves as descriptive prompts suitable for text-to-image generation, we utilized Stable Diffusion XL-1.0~\cite{stablediffusion} to convert all scripts into lists of images, facilitating evaluation by users based on the scenes.

\textbf{Metrics.} By assessing how different script generation methods affect the visual style and content of scene images, we compare the proposed script generation method with baselines. Specifically, we defined the following three metrics for a subjective evaluation:
\begin{itemize}
  \item \underline{\textit{Music Correspondence}}: how well the visual style (e.g. lighting, color) and music style (e.g. genre, emotion, and rhythm) are correlated. This metric estimates the usefulness of incorporating music captions (descriptions about genre, instrument, vocal, mood,
tempo, culture, and sonority features of a music clip) in script generation.
  \item \underline{\textit{Content Coherence}}: the content coherence of the scene images that are transformed from generated scripts. This metric estimates the usefulness of considering narrative coherence in script generation.
  \item \underline{\textit{Style Consistency}}: the style consistency of the scene images that are transformed from generated scripts. This metric estimates the effectiveness of incorporating style keywords into script generation.
  %   \item \underline{\textit{Content Coherence}}: the content coherence of the scene images that are generated based on the scene prompts within the same music. This metric estimates the usefulness of considering narrative coherence in script generation.
  % \item \underline{\textit{Style Consistency}}: the style consistency of the scene images that are generated based on the scene prompts within the same music video script. This metric estimates the usefulness of incorporating style keywords in script generation.
\end{itemize}

%We expected to assess the effectiveness of introducing music captions using "music correspondence", evaluate the impact of adding style keywords using "style consistency," and verify the narrative coherence of the script through "content coherence".

\textbf{Participants and Procedure.} A group of 20 participants (12 females, 8 males) aged 18-28 years old ($M=23.7, SD=2.32$) were recruited for the study. In the study, participants must compare and rate the scene images generated based on baselines and our script generation method via an online questionnaire. The order of the scene images in the questionnaire was randomized. While showing the scene images, the corresponding music was also played. The users were asked to make a rating after the music was finished based on a 5-point Likert scale (1="very poor", 5="very good"). 

%We distributed an online questionnaire to participants, where scenes transformed by scripts generated from three methods were grouped by music and presented randomly within each group. Participants were required to rate the scenes on a 5-point Likert scale (1="very poor", 5="very good") after listening to the music.

\begin{figure}
    \centering
    \includegraphics[width=0.7\textwidth]{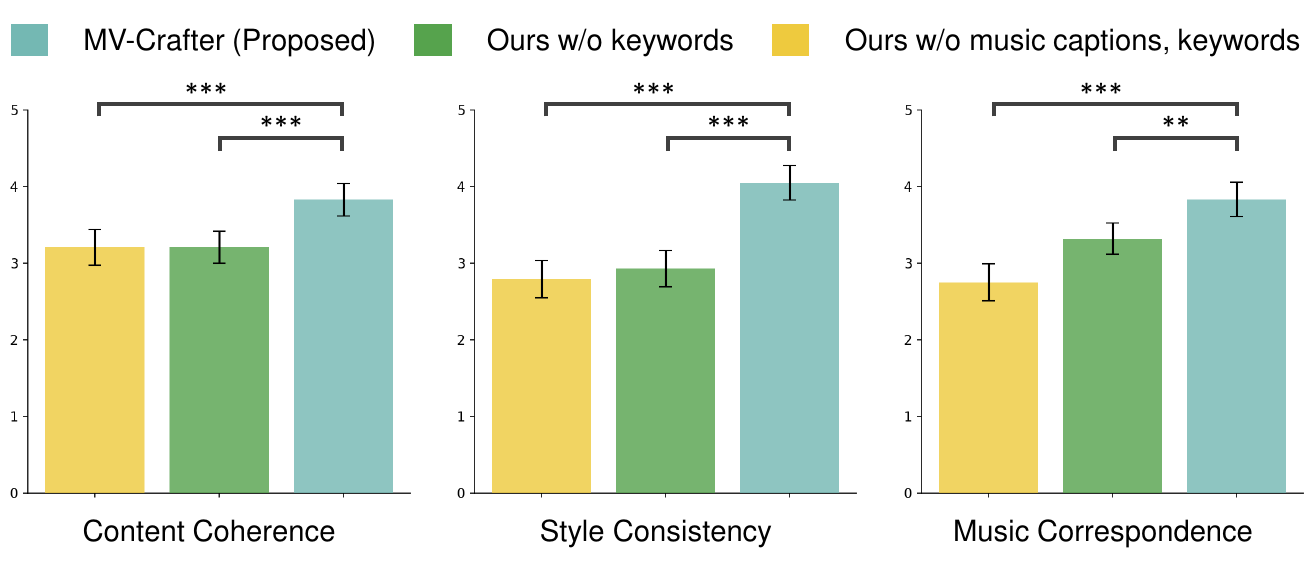}
    \caption{The results of ablation study on script generation module. Horizontal brackets indicate pairwise significant difference ($*: p<.05, **: p<.01, ***: p<.001$). The error bars represent the $95\%$ confidence intervals.}
    \label{fig:ablation-script}
\end{figure}

\textbf{Quantitative Results.}
We conducted one-way ANOVA and post-hoc Tukey tests on the questionnaire results to identify pairwise differences among $B_1$ (w/o style keywords), $B_2$ (w/o music captions, style keywords), and our script generation method. The analysis results are illustrated in Fig.~\ref{fig:ablation-script}. For content coherence, our method ($M=3.83, SD=1.08$) performed significantly better than the baselines, $B_1$ ($M=3.21, SD=1.04; p<.001$) and $B_2$ ($M=3.21, SD=1.17; p<.001$). For style consistency, our method ($M=4.05, SD=1.11$) was rated significantly higher than the baselines, $B_1$ ($M=2.93, SD=1.18; p<.001$) and $B_2$ ($M=2.79, SD=1.21; p<.001$). For music correspondence, our method ($M=3.83, SD=1.14$) significantly outperformed $B_1$ ($M=3.32, SD=1.02; p<.01$) and $B_2$ ($M=2.75, SD=1.20; p<.001$). 

In general, ablating style keywords ($B_1$) would result in an overall inconsistency in visual style, leading to a lack of coherence in the visual content. If both music captions and keywords were ablated ($B2$), the music correspondence would be significantly lower ($p<.01$) compared to the $B1$ method, and the video content would fail to integrate music information. These analyses demonstrated that the addition of keywords in the third step effectively ensures consistency in video style, while the introduction of music captions in the second step effectively incorporates music information into the video content. The designed three-step generation module can proficiently produce a script that is coherent in content, consistent in style, and matched with the music.

% We conducted one-way ANOVA and post-hoc Tukey tests on the questionnaire results to identify pairwise differences between methods and illustrated the analysis results in Fig~\ref{fig:ablation-script}. For content coherence, our method ($M=3.83, SD=1.08$) performed significantly better than the baselines, w/o keywords ($M=3.21, SD=1.04; p<.001$) and w/o music captions, keywords ($M=3.21, SD=1.17; p<.001$). For style consistency, our method ($M=4.05, SD=1.11$) was rated significantly higher than the baselines, w/o keywords ($M=2.93, SD=1.18; p<.001$) and w/o music captions, keywords ($M=2.79, SD=1.21; p<.001$). For music correspondence, our method ($M=3.83, SD=1.14$) significantly outperformed w/o keywords ($M=3.32, SD=1.02; p<.01$) and w/o music captions, keywords ($M=2.75, SD=1.20; p<.001$). 

% In general, ablating style keywords would result in an overall inconsistency in visual style, leading to a lack of coherence in the visual content. If both music captions and keywords were ablated, the music correspondence would be significantly lower ($p<.01$) compared to the only keywords ablated method, and the video content would fail to integrate music information. These analyses demonstrated that the addition of keywords in the third step effectively ensures consistency in video style, while the introduction of music captions in the second step effectively incorporates music information into the video content. The designed three-step generation module can proficiently produce a script that is coherent in content, consistent in style, and matched with the music.

\begin{figure*}
  \centering
  \includegraphics[width=\textwidth]{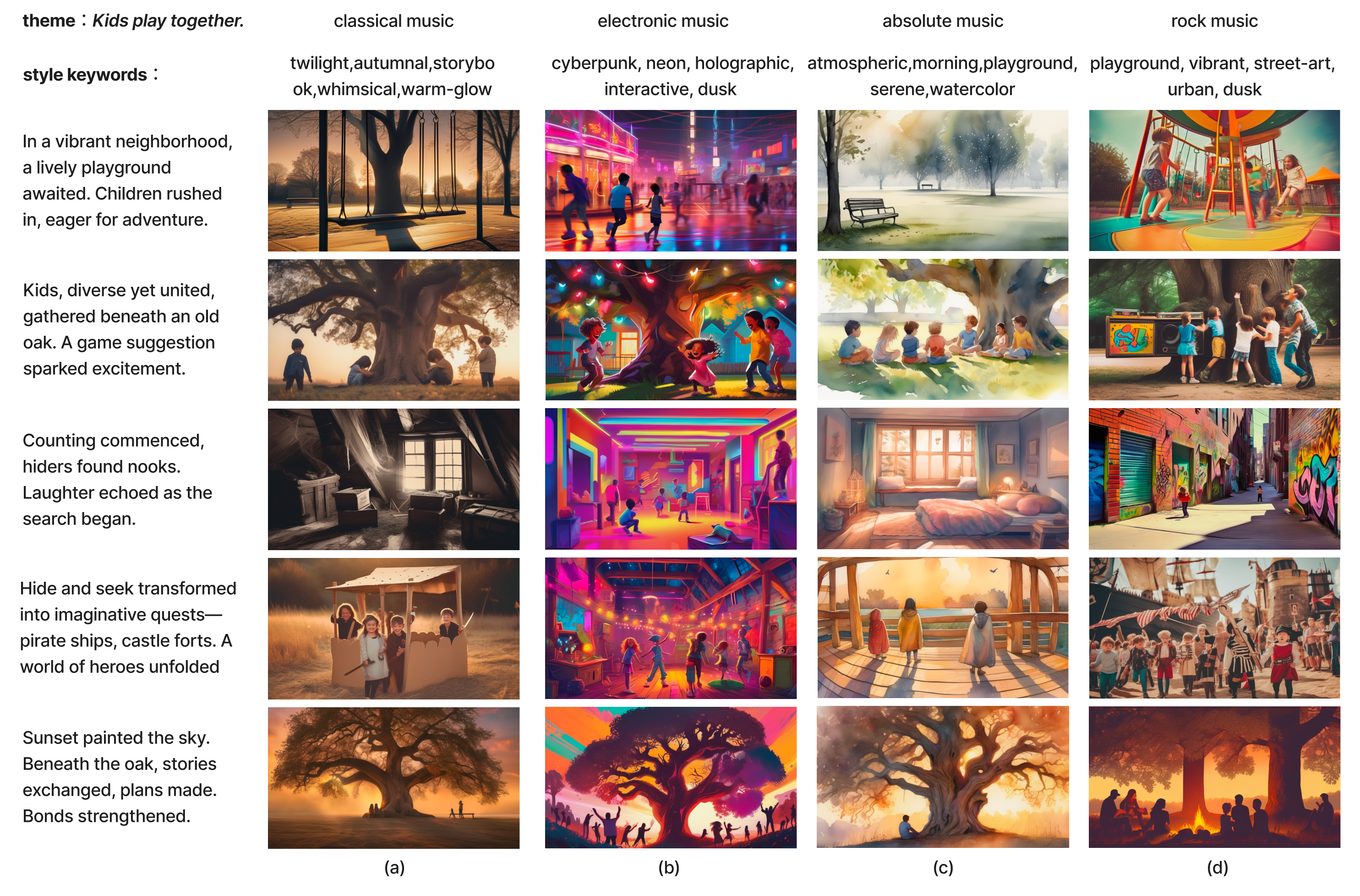}
  \caption{Qualitative results of script generation with music of different genres. We present the scenes generated by scripts produced from four input music, including classical music \textit{Canon in D} by Johann Pachelbel, electronic music \textit{Bullet Train} by Tubebackr, absolute music \textit{One Summer's Day} by Joe Hisaishi and rock music \textit{We Will Rock You} by Queen.}
  \label{fig:ablation-qualitative-1}
\end{figure*}

\paragraph{\bf Qualitative Evaluation.}
Fig.~\ref{fig:ablation-qualitative-1} illustrates the scene images generated by Stable Diffusion based on the music video scripts produced by our technique. All the scripts and images are generated based on the same input theme, i.e., \underline{\textit{``Kids play together"}} but for music pieces in different genres, thus resulting in images with completely different look-and-feels. For example, in terms of classical music Fig.~\ref{fig:ablation-qualitative-1}(a), our method generates style keywords such as \textit{"twilight, autumnal, storybook, whimsical, warm-glow"}, resulting in cozy and warm scene images.  In comparison, for electronic music Fig.~\ref{fig:ablation-qualitative-1}(b), our methods generate style keywords such as \textit{"cyberpunk, neon, and holographic"}, resulting in images with a cyberpunk style. Similarly, the scene images of absolute music (Fig.~\ref{fig:ablation-qualitative-1}(c)) and rock music (Fig.~\ref{fig:ablation-qualitative-1}(c)) also possess the right look-and-feel that matches their respective music genres. These results fully demonstrate the power of our script generation method.

\subsection{\highlight{System Usage Evaluation}}
\highlight{To further evaluate the usability of MV-Crafter system, we conducted a user study focused on assessing how effectively users interacted with the system and performed tasks related to music video generation.} 

\highlight{\textbf{Participants.} We recruited 9 participants (6 females, 3 males) aged 22 to 30 $(M=24.2, SD=2.7)$, denoted as P1-P9, through personal contacts and informal word-of-mouth within our network. Among these, 4 participants were novices, 3 were beginners, and 2 were competent in music video creation, ensuring a range of expertise levels for the study.} 

\highlight{\textbf{Procedure.} First, we collected participants' personal information through a questionnaire. Participants were then given a 5-minute introduction to the system's purpose and usage process. After confirming their understanding of all the features, they were asked to select one music track from the three provided and use the system to create a music video. During the creation process, participants were encouraged to think aloud. Upon completing the creation task, participants
were asked to rate their experiences with MV-Crafter using a post-study
questionnaire, which included the Creative Support Index and usability metrics rating on a 7-point Likert scale. To gain deeper insights, semi-structured interviews were performed to collect their feedback on
MV-Crafter. The whole process lasted for about one hour and was recorded for later analysis.}

\begin{figure}
    \centering
    \includegraphics[width=0.7\textwidth]{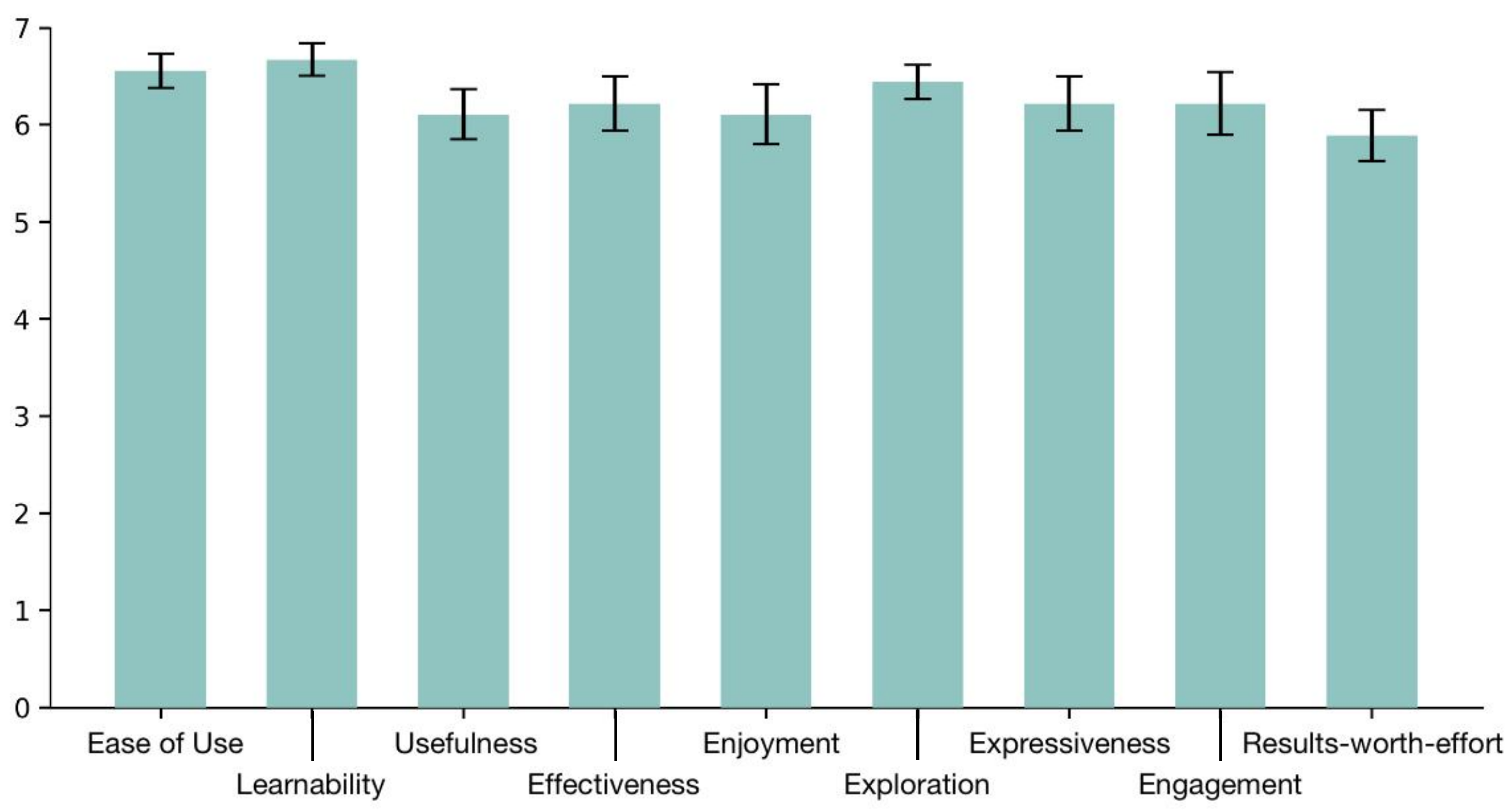}
    \caption{\highlight{The ratings of MV-Crafter from different measurements. The error bars represent the standard errors.}}
    \label{fig:system-rating}
\end{figure}

\highlight{\textbf{Results.} Fig.~\ref{fig:system-rating} shows the mean ratings of the participants' experiences with MV-Crafter. The system received positive feedback, particularly for learnability and ease-of-use, highlighting its user-friendliness. This was further supported by the interviews, where participants shared a wealth of positive comments. Below, we summarize the key insights from these interviews.}

\highlight{\underline{\textit{System Usability.}} All the participants found the MV-Crafter system to be intuitive and easy to use. Many noted that the interface was visually clear and simple (P1-P3, P6, P9), facilitating quick understanding of the workflow. The system's ability to allow for modifications, such as adjusting prompts and editing images, was also highlighted as a positive aspect. However, some participants suggested that more flexibility in rearranging the order of scenes and adjusting the scenes would improve the overall user experience. P5 suggested that \textit{"I hope the system allows for more control over the number and length of the scenes"}.}

\highlight{\underline{\textit{Creative Efficiency.}} Most participants appreciated how MV-Crafter facilitated rapid creative development by generating initial scripts and scenes. Many stated that this \textit{"significantly reduced the time spent brainstorming"} and \textit{"helped visualize their impressions of the music"}. P6 also said that \textit{"generating alternative scenes increased their creative options"}. Additionally, some participants felt the system could help them generate a prototype. P9 highlighted that the system offered a \textit{"0 to 0.7 effect"}, meaning it enabled beginners to start from scratch and generate a basic version of the music video, providing a valuable foundation when they had no clear direction to begin with.}

\highlight{\underline{\textit{Video Quality.}} In terms of video quality, participants were generally satisfied with the synchronization between the generated video and the music. All participants noted that the video movements and music beats were well synchronized, while some suggested that \textit{"enhancing the transitions between scenes could improve the overall flow"} (P3, P8). However, several participants pointed out issues with blurry scenes and a lack of visual coherence. They felt that improving the clarity and continuity between scenes would contribute to a more polished and professional video output.}

\section{Limitations and future work}
Despite the positive evaluation results indicating that MV-Crafter is promising to generate high-quality music videos, we would also
like to summarize and discuss several limitations that were found during
the design and implementation process and mentioned by our users. We hope to point out several potential future research directions by highlighting these limitations.

\textbf{Eliminating Motion Artifacts.} Some generated music videos exhibit motion artifacts. We find that these artifacts primarily stem from the video generation model, partly due to excessive interpolation during synchronization with longer music clips. Fortunately, the field of video generation is rapidly advancing with models like Sora~\cite{sora} capable of generating videos up to a minute in length with high visual quality. Our pipeline can be easily scaled up with the updated video generation models in the future.

% If there are more advanced open-source video generation models in the future capable of producing artifact-free, high-frame-rate, and longer-duration videos, we will replace the current video generation model to generate higher-quality music videos.

%\textbf{Improving Synchronization Quality.} The generated music videos do not exhibit a high level of synchronization as evaluated in the user study. Due to the low frame rate and limited motion range in AI-generated videos, extracting noticeable visual beats is challenging. With the flexibility of our pipeline, synchronization quality will notably improve as video generation models advance.

\textbf{Enhancing Narrative Coherence.} Although we have demonstrated that our approach achieves the best coherence compared to other AI-generated systems in the user study, \highlight{the narrative quality of some videos doesn't appear as strong as human-made productions.} We suspect that is due to the current system lacking consistency in character identity. To address this issue, future work should integrate adapters (e.g. ControlNet~\cite{zhang2023controlnet}, IP-Adapter~\cite{ye2023ipdapter}) into the video generation module and enhance the script generation module to generate character depictions.

% \textbf{Accelerating Generation Process.} The current system design and
% implementation have some performance bottlenecks. In particular, it
% usually takes approximately 30 minutes to generate a 1-minute music video. However, this is mainly due to the performance limitation of nowadays video generation techniques.It usually takes around 1 minute to generate a video clip of 2-4 seconds due to the high computational cost. Our experiment showed that both the script generation and audio-video synchronization only take a few seconds to get the output.

\textbf{Accelerating Generation Process.} The current system design and
implementation have some performance bottlenecks. In particular, it
usually takes about 30 minutes to generate a 1-minute music video. However, this is mainly due to the performance limitation of nowadays video generation techniques. Generating a video clip of 2-4 seconds typically requires approximately 1 minute due to the high computational cost. It is anticipated that the generation time will be significantly reduced as video generation models advance.

% As maintaining consistent character identity in long video generation is a remaining unsolved problem, we leave it as future work. Adding more user controls across the system to facilitate user adjustments is also a promising direction we would explore.

\highlight{\textbf{Improving Control Flexibility.} Several participants (P4-P6, P8) expressed the need for more flexibility in rearranging the scene order and adding transition effects between scenes. P1 also suggested that the system could provide more style keywords for users to choose from during generating scripts. Future work should explore expanding the system’s editing features to provide users with more creative control and enhance their overall experience.}

\section{Conclusion}
In this work, we present a novel interactive system empowering non-professionals to create high-quality, rhythmic music videos through intuitive interactions. Our framework takes advantage of LLM and diffusion models for generating scripts and video clips. We propose a three-step script generation method that incorporates musical semantics when generating scripts. We introduce a novel dynamic synchronization method tailored for music videos to achieve alignment between music and videos of arbitrary lengths. Extensive experiments have validated the system's effectiveness in music video generation.

\begin{acks}
Nan Cao is the corresponding author. This work was supported by the National Key Research and Development Program of China (2023YFB3107100), NSFC 62072338, NSFC 62372327, and NSFC 62061136003.
\end{acks}

\bibliographystyle{ACM-Reference-Format}
\bibliography{bibliography}

\clearpage
\appendix

\section{Case study of script generation}

\begin{figure*}[h!]
  \centering
  \includegraphics[width=\textwidth]{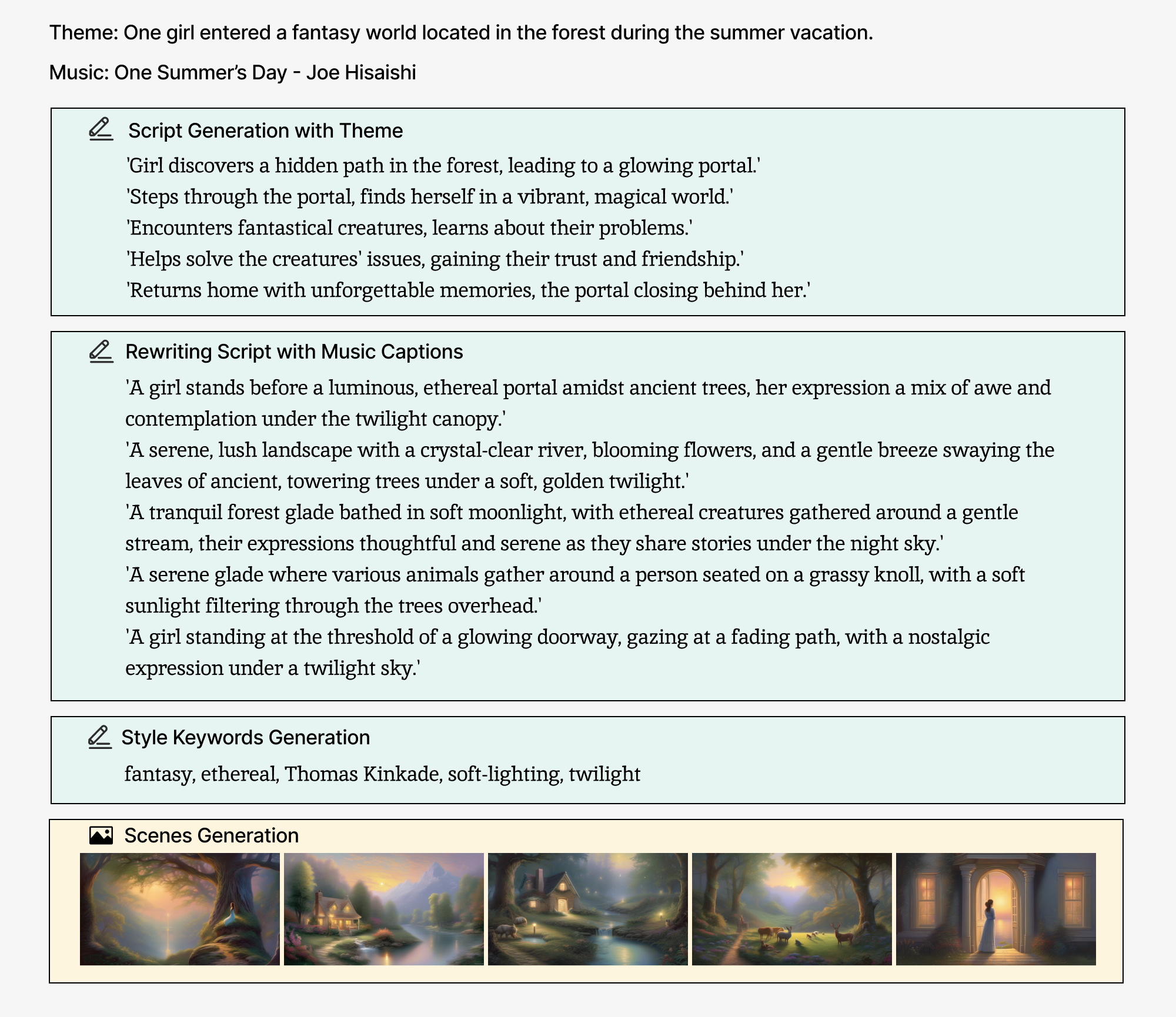}
  \caption{The complete script generated by the script generation module in the case study.}
  \label{fig:script-example}
\end{figure*}

\end{document}